%% file: stelpops.tex
\documentclass{report}

\usepackage{graphicx} 
\usepackage{multirow}
\usepackage{array} 
\usepackage{natbib}

\begin{document}
\input chapter_peletier/stellarpopulations

\end{document}

%% file: chapter_peletier/stellarpopulations.tex
\input chapter_peletier/bmacros

\setcounter{chapter}{1}
\pagestyle{myheadings}
\markright{Reynier Peletier\hfil Stellar Populations\hfil}

\author{Reynier Peletier\\ (Kapteyn Astronomical Institute)\\  
         University of Groningen,\\ Postbus 800, 9700 AV Groningen,NL \\ 
	(To appear in XXIII Canary islands winter school of astrophysics\\
	ed.\ J. Falc\'on-Barroso \& J.H. Knapen)}

\title{Stellar Populations}
\maketitle

\abstract {This is a summary of my lectures during the 2011 IAC Winter School in
Puerto de la Cruz. I give an introduction to the field of stellar populations in
galaxies, and highlight some new results. Since the title of the Winter School
was {\it Secular Evolution in Galaxies} I mostly concentrate on nearby galaxies,
which are best suited to study this theme. Of course, the understanding of stellar
populations is intimately connected to understanding the formation and evolution
of galaxies, one of the great outstanding problems of astronomy. We are
currently in a situation where very large observational advances have been made in
recent years. Galaxies have been detected up to a redshift of 10. A huge effort
has to be made so that stellar population theory can catch up with
observations.  Since most galaxies are far away, information about them has to
come from stellar population synthesis of integrated light. Here I will discuss
how stellar evolution theory, together with observations in our Milky Way and
Local Group, are used as building blocks to analyze these integrated stellar
populations.}

\tableofcontents

\section{Introduction}\label{Introduction}

We are living in an era in which we are able to see the light of galaxies at or
close to redshift 10. Many galaxies above a redshift of 3 are already
known, and we are starting to discover their clustering properties. These
galaxies must undergo an evolution which leads to the galaxies we see around us,
of which we have catalogues containing millions of individuals. Understanding
how this evolution has taken place is a major task that the current generation
of astronomers will have to address. A major tool needed to study galaxy
evolution is stellar population synthesis. At high redshift galaxy are not more
than point sources of which we can measure the flux in one or more bands, or a
spectrum, if we are lucky. The information contained in such a spectrum can, however,
be very powerful. It can tell us how old the stars are that emit the light that
we see, and give us some clues about their internal composition. This
information, together with spatial information about the distribution of
galaxies, and information about the gas content of these objects, is crucial
input for large galaxy formation simulations, which try to simulate the
formation and evolution of galaxies in the Universe. 

To interpret the fluxes and spectra of high redshift galaxies, we need models that
tell us how stellar populations evolve as a function of time. With them, we can
then use the fact that astronomers can look back in time, to determine a 3D
movie of the evolution of galaxies in the Universe. Even some predictions can be
made about the future. Such models are best made using the detailed information
around us, from the Sun, which shows us in detail how one particular star evolves,
to other stars, to galaxies in the Local Group, which we can resolve into
individual stars, to galaxies in the local Supercluster, for which we can obtain
important spatial information. Stellar population models have been built on knowledge from
many different fields, and are continuously improving in quality, as our
understanding of many physical processes improves, the capacity of our computers
increases, and our instruments and telescopes become larger and better. 

Research on stellar populations is relatively old. In the beginning of the 20th
century the Hertzsprung-Russell diagram was discovered, i.e., the relation between the
color and luminosity of stars. The HR diagram showed that stars evolve in an ordered
way. Shapley (1918) could show from Cepheids that globular clusters are large systems of stars at several kpc from us centered around a point which
later became known as the center of our Galaxy. Soon afterwards, it was found
that nebulae such as M31 were even larger systems than globular clusters (galaxies) much further away.
It was realized by Baade (1944) that some regions of M31 were similar to our Galactic
disk as far as the kind of stars present were concerned, and that the bulge of
M31 looked like our Galactic halo. As a result, one realized that the variation
of properties along Hubble's galaxy classification sequence was strongly
connected with star formation in galaxies. Early works studying galaxy colors
to derive information about their stellar contents were by Stebbins, Whitford,
Holmberg, Humason, Mayall, Sandage, Morgan, and de Vaucouleurs (see Whitford
1975 for a summary of the early work in this field). A nice review about the
known galaxy properties is given by Roberts (1963). Roberts' analysis
established the basic elements of our current picture of the Hubble sequence as
a monotonic sequence in present-day SFRs and past star formation histories.

To go further than recognizing differences between galaxies, and to be able to
give quantitative predictions for stellar populations in galaxies that can be
tested with real data, a paradigm of evolutionary stellar population models was
developed by Tinsley (1968, 1980, etc.). In Kennicutt (1999) the importance of
Tinsley's stellar population work has been summarized. She developed a
methodology for constructing evolutionary synthesis models for colors, gas
contents and chemical abundances with time, heavily influenced by the ideas
about galaxy formation published in 1963 by Eggen, Lynden-Bell \& Sandage
(1962). This {\it evolutionary synthesis} paradigm, which is still being used by
our current, most sophisticated models, needs input from stellar evolution
calculations, nucleosynthesis models, and cosmological models, and is made to
fit the observed colors and magnitudes of stars and galaxies, as well as their
line strengths. An important result of her models was that she was able to
describe the evolution of colors and magnitudes along the Hubble Sequence as a
function of one parameter, which could be M/L ratio, gas fraction or age. A
second result was that the luminosity evolution of passive galaxies was much
larger than had been estimated earlier. Before her work it was claimed that  one
could constrain q0 and $\Lambda$ from the evolution of magnitudes  and colors.
However, nowadays one knows that the evolution due to  aging and other
population effects is so large compared to effects of cosmological signatures
that this method is not very useful. Because she was able to indicate the
importance of stellar population research, Tinsley became the focal point of
galaxy evolution research and a role model for younger scientists. The
conclusion from her work was that galaxy evolution was an observable phenomenon
(see Kennicutt 1999).

Tinsley's ideas were incorporated into the models of Faber (1972), and of
Searle, Sargent \& Bagnulo (1973), who also added a parameterization for the
Star Formation History. For a number of years people seriously tried to compete
with there evolutionary models using {\it empirical} models that did not need
any input from theory (e.g. O'Connell 1980, Bica 1988), but these efforts have
died out, since it is hard to learn much about the evolution of stellar
populations in this way. 

Since the time of Tinsley, our observing capacity has improved enormously. We
have CCDs, linear devices that make it possible to record accurately images and
spectra of galaxies. Telescopes have increased in size, but, more importantly,
the Hubble Space Telescopes has caused a revolution in the field of stellar
populations. The high spatial resolution, the long integrations, and the
changing paradigm to work with sometime very large, surveys, has made it possible
to obtain high S/N data of very far away galaxies, and at the same time very
high quality, detailed images of nearby objects. Such images become even more
powerful in the presence of the large SDSS survey, which makes it possible to
carefully select galaxies for detailed studies using HST.  Also very important,
compared to Tinsley's time, is that we have started to explore most of the
electromagnetic spectrum, and also non-baryonic matter. Knowing about the huge
presence of dark matter completely changes our idea about the expansion of the
Universe and the interactions between galaxies, and is therefore crucial to
understand galaxy evolution. 

\begin{figure}
\begin{center}
\includegraphics[width=\textwidth,clip]{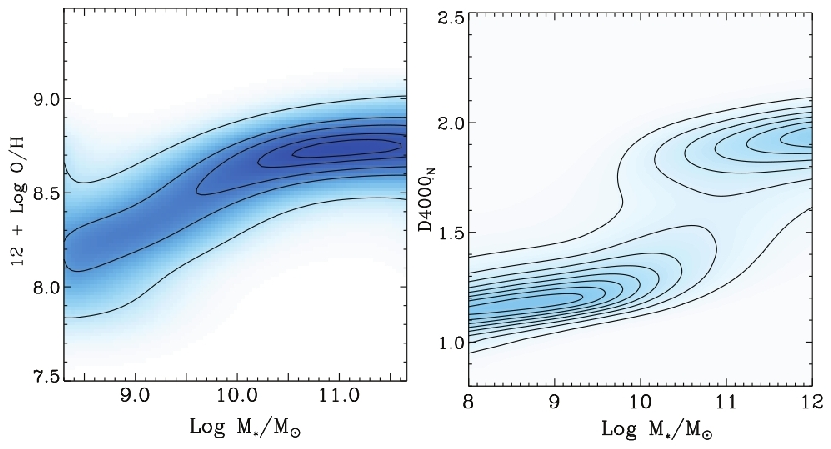} \includegraphics[width=0.8\textwidth,clip]{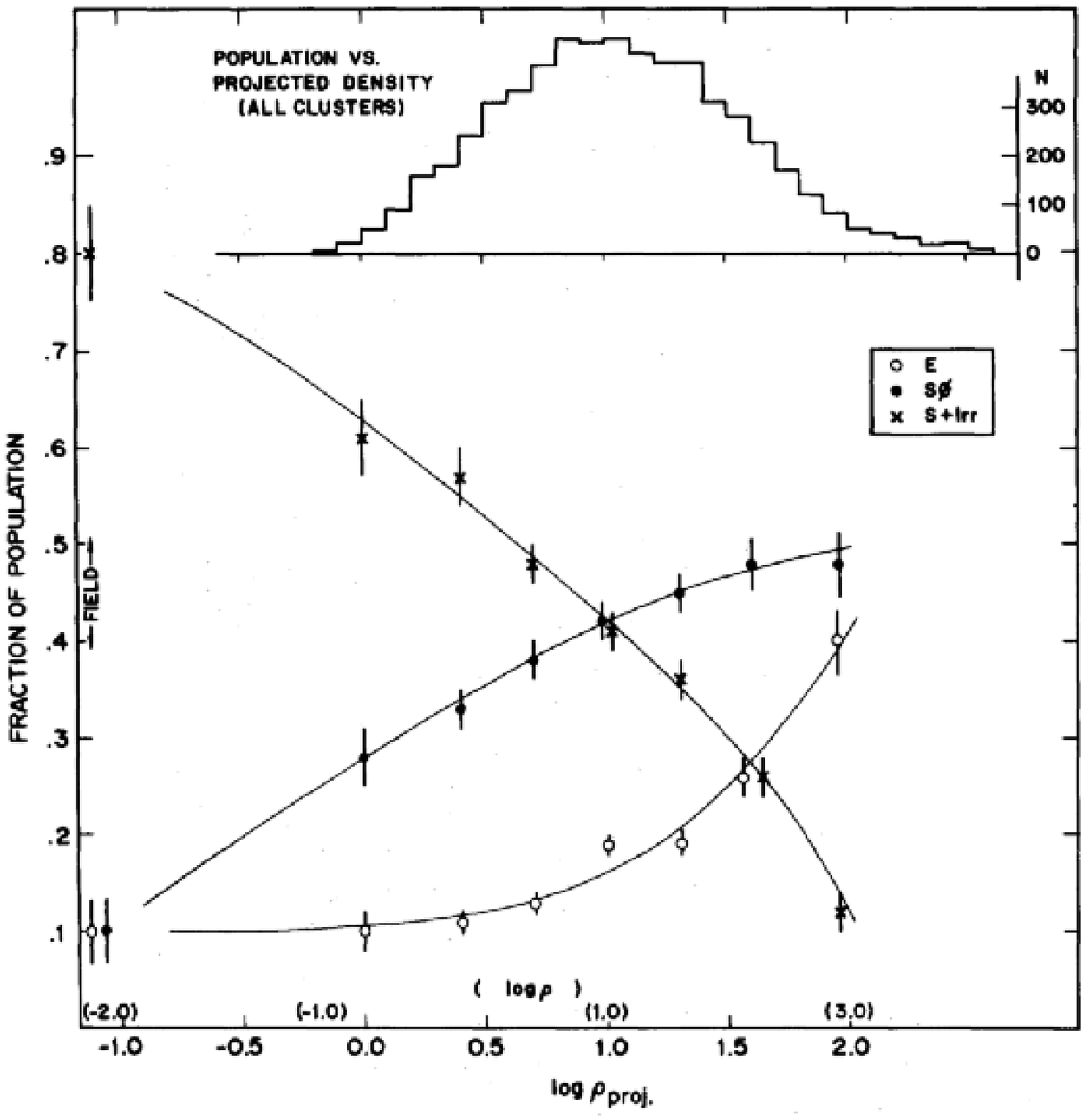} \caption{On the top left is shown the
mass-metallicity relation for SDSS DR6 (Tremonti et al. 2004). Top right: The relation between stellar mass and the
4000\AA\ break in the SDSS DR7 (Kauffmann et al. (2003)). These figures have been reproduced from Brinchmann (2010).
Bottom panel: On the left the Morphology-Density relation
(Dressler 1980), showing the strong relation between  the galaxy density and the fraction of early-type galaxies. }
\label{Brinchmann}
\end{center}
\end{figure}

Fig.~\ref{Brinchmann}, from Brinchmann (2010) indicates some very basic
problems in our field, which will hopefully be solved in the coming decades.
On the left the morphology-density relation
by Dressler (1980). Understanding why in clusters in the local Universe the
fraction of early-type galaxies, both dwarfs and giants, is so much larger than
in the field, is a fundamental question, which needs full attention of our
community. Then in the middle the mass metallicity relation. This relation is one of the most
fundamental relations in  the field of galaxy formation and evolution. The
solution depends on how a galaxy is formed from the primordial gas, on feedback
at all times, on interaction with its environment, and on many other things. And on
the left the dichotomy in the mass -- D$_{4000}$ relation, which loosely
can be called the mass -- age relation. It is clear that smaller galaxies form
their stars later. Why is this, and why is there a distinct break?  

The contents of this review will be the following. I will start with a quick look at galaxy images, so that
the reader can get a feeling of the different stellar populations (Section \ref{quicklook}). After that I
will discuss various aspects of Stellar Population Models (Section \ref{models}), such as stellar evolution
theory, stellar libraries, the Initial Mass Function and abundance ratios. In the next Section
(\ref{indgal} I will discuss stellar population analysis of individual galaxies, using colors and spectra,
in the optical and elsewhere. I will discuss obtaining star formation histories, and end this section with
a few words on surface brightness fluctuations. Following this, in Section \ref{scalrel} I will discuss stellar population
analysis using scaling relations with ensembles of galaxies. In Section \ref{gradients} radial gradients in the galaxies
will be discussed. I will end with some prospects for the future.

\section{A quick look at galaxy images}\label{quicklook}

In 1943, aided by the blackout in Pasadena caused by the second world war, Baade for the first time was
able to resolve M31 and its companions into individual stars (Baade 1944). Realizing that the stellar
populations in NGC 205 and M~32 are similar to the stars in globular clusters in our Galaxy, Baade
recognized that stellar populations fall into 2 groups: Population I stars, which look like the stars
in the solar neighborhood, and Population II stars, which are similar to those in globular clusters.
Characteristic for Population I are high luminosity O and B stars, as well as open clusters. Characteristic
for Population II are short-period Cepheids. Early-type galaxies, according to him, only contained
Population II, while in late and intermediate-type galaxies a mix of both populations was found.   Only
much later on (in the 1970 and 1980's by Whitford (1978), Whitford \& Rich (1983), Rich (1988), Terndrup
(1988), etc. ) it was found that these stars were very different, with bulge stars being metal rich, while
halo stars are very metal poor. Both populations, on the other hand, have in common that they do not
contain any blue giant stars. 

At present, we think that our Galaxy contains several components: the rotating (thin) disk with population
I stars, the thick disk (Gilmore \& Reid 1983), the stellar halo and the Galactic bulge. In the halo
several globular clusters are found. Some others have properties more like the thin disk and are therefore
called  disk globular clusters (Zinn \& West 1984).

\begin{figure}
\begin{center}   
\includegraphics[width=\textwidth,clip]{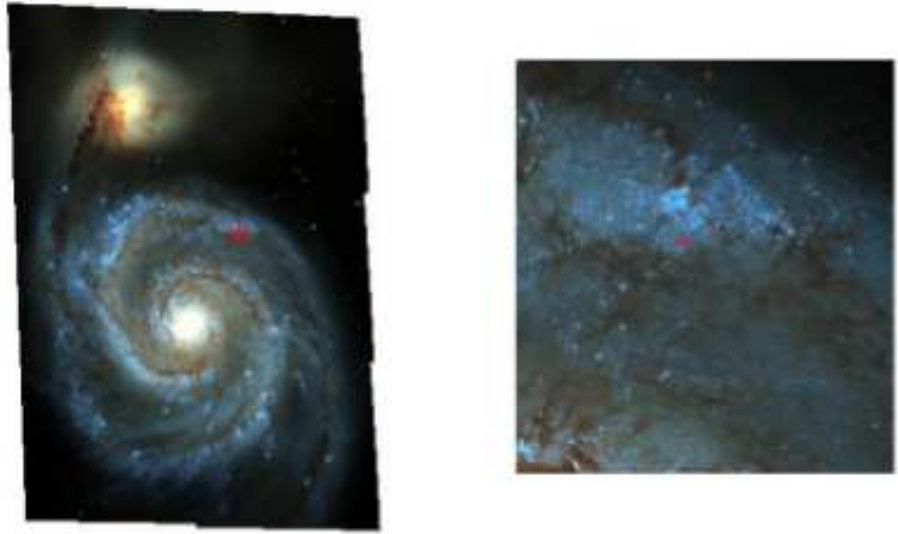}
 \caption{Left: Composite HST F814W/F555W/F435W image of M~51 (from the Hubble Legacy Archive). Right: Zoom at the position of the cursor, showing the stellar populations around a spiral arm. }
\label{M51a}
\end{center}
\end{figure}

\begin{figure}
\begin{center}   
\includegraphics[width=\textwidth,clip]{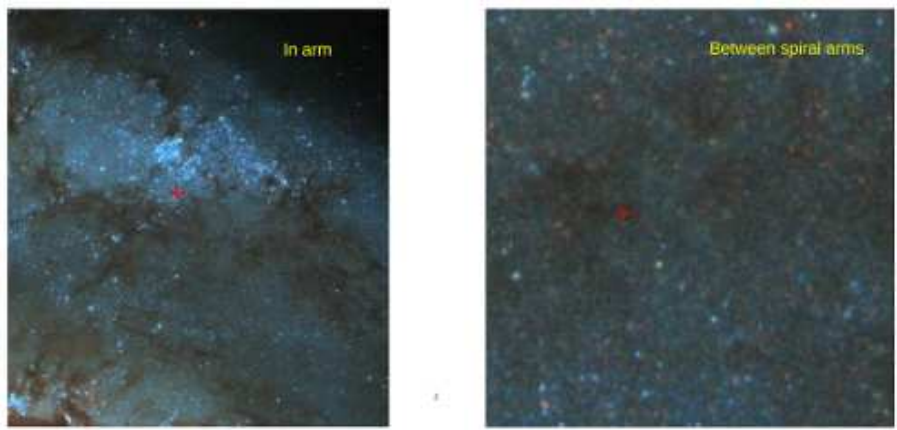}
 \caption{Magnifications of Fig.~\ref{M51a} in (left) and out of the spiral arm (right). Note the enormous difference in resolved stellar populations. }
\label{M51b}
\end{center}
\end{figure}

We are in the lucky situation that now with the Hubble Space Telescope we can imitate Baade's observations
for galaxies up to about 10 Mpc, i.e. a factor 50 further away. The Hubble Legacy Archive ({\tt
http://hla.stsci.edu/}) offers fantastic images enabling everybody to {\it observe} galaxies at a
resolution which is a factor 10 better than one can get from the ground. As was the case for me, when I was
an undergraduate student in Leiden in the 1980's when I studied the Palomar Sky plates: it is extremely
instructive to look at images of the Universe. For example, by simply studying maps of galaxy clusters such
as Virgo or Coma one gets a very good idea of the morphology-density relation. I will show here the galaxy
M51 at 8 Mpc. Taking the composite F814W/F555W/F435W image one clearly sees the spiral arms, the dust
lanes, the position of star forming regions close to the dust lanes, etc. (Fig.~\ref{M51a}).  Zooming in
into the area marked with a red cross one sees a large star forming complex  in the spiral arm
(Fig.~\ref{M51a}(right)), together with many small and large dust lanes. Zooming in one step further into
the star forming complex (Fig.~\ref{M51b}(left)) one sees the red and blue (super)giants individually
resolved. The stellar populations here are similar to a young open cluster, containing both young and old
stars. Outside the spiral arm (Fig.~\ref{M51b}(right)) fewer stars are being seen. The brightest stars have
disappeared, both the blue and the red ones, and the stellar population appears to be much more uniform
that in the spiral arm. Here one barely resolves the brightest red giants of a stellar population, which
looks similar to Population II. 

A factor of 2 in distance further from us, we can still resolve individual stars with the Hubble Space
Telescope. In the Coma cluster, however, another factor of 6 further away, this is not possible any more.
Here information about individual stars can still be obtained using surface brightness fluctuations (see
subsection \ref{sbf}), analyzing the statistics of the light in individual pixels. When going even further
away the only way to derive information about stellar populations in galaxies is by analyzing the
integrated light. In most of this paper we will discuss ways to do this, and the results that are
obtained.

\section{Stellar population models and their ingredients}\label{models}

With stellar population synthesis one wants to determine as much information from the stellar populations
of an unresolved galaxy from a single spectrum as possible.  Having seen how complicated our Milky Way is,
it is clear that one cannot retrieve all the information that one would like to have. The integrated light
contains contributions from stars with a large range in metallicity, mass and age, and even for stars of a
fixed metallicity the relative distribution of elements might be different from one star to another. Since
all this together is too much to obtain from a single spectrum, one often assumes an {\it Initial Mass
Function (IMF)}, the number of stars per mass interval at the moment that the stars were born, and tries to
derive the star formation history (SFH) and the metallicity distribution of the stars in a galaxy.

The spectrum of an integrated stellar population is obtained by integrating all stars along the line of
sight. 

\begin{equation} S_{\lambda}(t,Z) = \int_{m_{low})}^{m_{up}} S_{\lambda}(m,t,Z)~N(m)~F(m,t,Z)~dm
\end{equation}  

Here $S_{\lambda}(m,t,Z)$ is the spectral energy distribution (SED) of a star of mass $m$ (going from
$m_{\rm up}$ to $m_{\rm low}$), age $t$ and
metallicity $Z$, and $F(m,t,z)$ the flux in a normalizing band. The total galaxy spectrum is then obtained
by adding the contributions of the various ages and metallicities. 

Modern stellar population synthesis assumes that the stellar populations in galaxies consist of a sum of
{\it Single Stellar Populations (SSP)}, building blocks of more complex stellar populations, 
entities that consist of all stars born at the same time, with the
same metallicity. For example, galaxies do not consist only of A-stars of 2 solar mass, but stars of the
full range of masses in the range of the IMF must exist. With this strong assumption, one can use stellar
evolution theory to calculate the integrated properties of these SSPs, and then decompose the SED of
galaxy into various SSPs.  Justification of the SSP approach dates back from the time of Tinsley, who
realized that clusters, both globular and open, could all be represented as single stellar populations.
When looking at the nearest galaxies, the SSP approximation remains valid (Section \ref{colouranalysis}). A
nice overview of evolutionary stellar population models is given by Maraston (2005). Current SSP models 
predict full spectral energy distributions (SEDs) by integrating stellar spectra along theoretical
isochrones. These spectra are selected from a stellar library,  that consist of observations of real stars
or of artificial spectra calculated using model atmospheres. Some of the presently most popular stellar
population models are those by Worthey (1994), Maraston (2005), Bruzual \& Charlot (2003) and later
unpublished versions, Le Borgne et al. 2004 (P\'egase HR), Leitherer et al. 2010 (Starburst99), Vazdekis et
al. (2010), Gonz\'alez Delgado et al. (2005) and Schiavon (2007).

Despite important progress in the last 2 decades, it is still not straightforward to routinely derive star
formation histories and metallicity distributions from spectra. Here I will briefly indicate a number of
reasons why this is the case, and in the coming sections I will elaborate on them. First of all, there is our
relatively poor understanding of the advanced phases of stellar evolution, such as the supergiant phase and
the asymptotic-giant-branch (AGB) phase. Stars in these phases are bright and short lived, so that they do
contribute significantly to the spectrum of galaxies, but their numbers in color-magnitude diagrams of
globular and open clusters are  not large enough to constrain the fraction of light of these populations in
the total spectrum. At the same time, mass loss, which is difficult to model, and is crucial for
calculating their evolutionary path, plays such an important role in these stars, that it causes
considerable uncertainties. Stars on the AGB contribute especially in the infrared, and in this region
especially these uncertainties are most important. 

A second, also very important problem is the age-metallicity degeneracy. In an integrated stellar
population, the main effect of increasing the metallicity is raising the opacities of stars on the red
giant branch (RGB), causing it to become redder, which implies that the integrated spectrum is redder, and
that most strong atomic features in the optical are stronger, since they are stronger in cooler stars. The
main effect of increasing the age is the same: the reddening of the RGB. This means that one can often not
derive both age and metallicity at the same time: the so-called age-metallicity degeneracy. This degeneracy
in fact can get worse: if there is dust in the galaxy, it will cause colors to become redder, acting in
the same way as increasing metallicity or age. However, only colors are strongly affected by the effects
of dust extinction. Line strengths are mostly not affected by extinction, except if blue and red continuum
are lying very far away from each other. As a result, ages and metallicities  are strongly degenerate,
although there are ways to measure both parameters separately.

A third problem is that abundance ratios of galaxies are not always the same as in the solar neighborhood.
This shows up in our Galaxy as a tight relation of [Mg/Fe] and metallicity [Fe/H] (Edvardsson et al. 1993).
It is thought that element formation is mainly taking place in supernovae of type Ia and II. In explosions
of SN type II (massive stars) the amount of O and other $\alpha$-elements, as compared to Fe, that is
produced is much larger than in SN of type Ia. It is believed that during the formation of halo and bulge
the number of SN type II relative to type Ia was much higher than during the formation of the disk, and as
a result the ratio of Mg/Fe abundance is higher in our Galactic halo and bulge than in the Galactic disk.
Not only does the fraction of $\alpha$-elements relative to Fe vary. Variations are also seen in the ratio of other elements to Fe (see the review of Henry \& Worthey 1999). At present stellar population models (both the stellar
isochrones and the stellar libraries) rarely contain abundance distribution other than solar. This means that
several features in observed galaxy spectra are not well fit by the best currently available stellar
population models.

\subsection{Stellar evolution theory}\label{evolution}

The basic theory of stellar interiors is one of the eldest in astronomy. Already Eddington (1926) was able
to somehow model the at that time just discovered Hertzsprung-Russel diagram with the differential
equations for stellar interiors. Currently all models are in principle based on his theory, although they
have been improved and expanded to take into account our growing understanding of the physics of stars. For
details about the current stellar evolution models, see e.g. the book of Salaris \& Cassisi (2006) on the
Evolution of Stars and Stellar Populations. Many ideas in this book are reflected in the website of the
BaSTI models (http://albione.oa-teramo.inaf.it/), by the previous authors and others.   One might wonder
why there is such a large variety of stellar models presented on this website. The reason is obvious: the
input physics is so uncertain that the final fit to the data will have to show what choices should be
made.  The stellar evolution model traces the evolution of stars of given mass and chemical composition
through the various evolutionary phases defined from the color-magnitude diagrams of open and globular
clusters, and provides the basic stellar parameters - bolometric luminosities, effective temperatures and
surface gravities  as a function of the evolutionary time. Apart from the BaSTI models there are several
other models available in the literature: e.g. the ones of Girardi et al. (2000), Marigo et al. (2008), Yi
et al. (2003), Lejeune \& Schaerer (2001), Dotter et al. (2007). 

As far as the main stages of stellar evolution, it does not really matter very much which models one takes
(see an earlier comparison study by Charlot, Worthey \& Bressan 1996). What matters are the {\it later}
evolutionary phases. For example, in the models of Vandenberg et al. (2000) no overshooting is used. The
models do not go beyond the RGB, since the authors do not trust our understanding of the physics beyond
that phase. It has to do with the treatment of convection and severe mass loss. The treatment of these
phases (mainly the HB and the TP-AGB) varies considerably between groups. The BaSTI models, for example,
give 2 sets of models for different mass loss rates $\eta$. For old stellar populations, for which the
TP-AGB phase is unimportant, this means that this uncertainty plays a smaller role (see Mouhcine \& Lan\c con 2002. In this paper it is shown that the fractional luminosity of the AGB in $J$, $H$ and $K$ is the highest between log(age) = 8.5 and 9.2, and reaches appr. 40\% in $J$, 50\% in $H$ and 60\% in $K$. The contribution also goes up with metallicity.). In the future larger telescopes will provide us color-magnitude diagrams down to lower
magnitudes, and the improved statistics obtained in this way will probably cause the stellar evolution
calculations for these phases to considerably improve. 

Another source of uncertainty is the He contents of stars (indicated as mass fraction Y). He comes mainly
from Big Bang Nucleosynthesis, but it strongly affected by H and He-burning in stars. An important problem
is that there are no easy ways to determine the He-content in stars. Atomic transitions generally require
high-energy photons, so are mostly found in hot stars, and are found in the blue or difficult to access UV.
Since He does affect the location of the stellar evolution tracks (see e.g. the BaSTI models) this is an
important problem. More generally, recent models assume that the helium fraction is related to the
metallicity Z (e.g. Vazdekis et al. (2010) use  Y ~=~ 0.23 + 2.25~Z).

Not only is there a problem with He, there are more than 100 other elements that each can have their own
relative abundance w.r.t. Fe. Traditionally, these are parameterized with Z, the mass fraction of all
elements other than H and He. Luckily, the relative abundances of these {\it metals} do not vary very much.
This is because it is thought that element production is a fairly uniform process: it is done mostly in
supernovae, but also in the envelopes of giant or supergiant stars. However, one knows that different types
of supernovae produce different relative fractions of elements: for example, since the rate of element
production depends on stellar mass, the relative distribution of elements ejected into the ISM in a SN
explosion must also depend on mass. Element ratios affect the stellar population models in 2 ways: both the
evolutionary tracks change position (see e.g. the BaSTI models), but more importantly, their line strengths
change more directly (see e.g. Lee et al. 2009). In section \ref{abunrat} this item will be discussed in
more detail. 

A third problem is the IMF (see before), which is very difficult to measure. This point will be discussed
in section \ref{IMF} and section \ref{IMFMass}.

Other problems occur when converting theoretical parameters (T$_{eff}$, g and abundances) to observational
ones, such as colors or spectra. These problems are enormous. Although it seems straightforward to
determine temperatures from spectra of observed stars, this is difficult for M-stars, which have such broad
absorption lines that complicated procedures are required to determine them (e.g. Lan\c{c}on et al. 2008).
For most stars reasonably accurate gravities are available from distances and CM-diagram fitting, and they
will become better when the results of GAIA are available. Determining abundances of stars is a complicated
subject, which is mostly done comparing line strengths of isolated transitions with stellar atmosphere
models (e.g. MARCS (Gustafsson et al. 2008)). See e.g. Ryde et al. 2004 for a review.  In the ideal situation these isolated
transitions are available. However, in practise for many elements no lines are available. Often, the
resolution of the spectra is so low that one a few elements can be investigated, or that less accurate
methods will have to be used, such as only determining a global metallicity for galaxies. For hot stars,
such as O or B stars, determining abundances is particularly difficult, since they contain almost no
lines. 

Apart from the models described above, the BaSTI website also provides models for pre main sequence stars
and for white dwarfs. The role of contact binary stars is not considered.  One way for them of manifesting
themselves is through Blue Straggler Stars (BSS), hydrogen-burning stars made hotter and more luminous by
accretion and now residing in a region of the color-magnitude diagram normally occupied by much younger
stars (e.g. Bond \& MacConnell 1971). The fact that they have been found in globular clusters (Piotto et
al. 2004) shows that their presence can be significant in galaxies. Recently, such blue stragglers have
also been detected in the Galactic Bulge (Clarkson et al. 2011). The reason that historically not much
attention has been given to BSS in stellar population models is that in color-magnitude (CM) diagrams Blue
Straggler Stars are not found in large quantities, and that integrated models of old ages do fit elliptical
galaxies. Since ignoring Blue Stragglers would cause elliptical galaxies to appear younger, it is
worthwhile to study this issue in the future.

\subsection{Stellar libraries}\label{libraries}

An important ingredient in stellar population models is the {\it stellar library}. The stellar library is essential
for the models. Its spectral resolution and wavelength range determine the spectral resolution and wavelength range
of the models. Its range in stellar types determines the applicability of the models, and to a large extent their
quality. It is defined as a set of spectra of stars, covering a range in stellar parameters, i.e. effective
temperature $T_{eff}$, gravity $g$, and metallicity [Fe/H]. More recently, also abundance ratios, such as [Mg/Fe] are
sometimes included (e.g. Milone et al. 2011). Stellar libraries can be theoretical (e.g. Hauschildt et al. (1997), Coelho et al. 2007) or
empirical. Many of them can be found in the compilation of D. Montes (http://www.ucm.es/info/Astrof/invest/actividad/spectra.html). A good, but somewhat outdated review
about empirical and theoretical stellar libraries is given by Coelho (2009). As mentioned in the section above, SSP
model spectra are determined integrating the spectral contributions from stars along isochrones, and weighting them
with the relative fraction of stars at every place along those isochrones. The required spectra here are obtained by
selecting the star (or linear combination of a few stars) with the required temperature, gravity and metallicity.
This is generally done using some kind of interpolation. 

The first often used stellar library was the LICK-IDS library (Faber et al. 1985, Burstein et al. 1984, Gorgas et al.
1993, Worthey et al. 1994). This library consists of 425 spectra obtained with a non-linear Image Dissector
Scanner, which each had been reduced to 25 low resolution line indices defined as equivalent widths of strong
absorption lines in the spectrum. For large galaxies, for which the velocity dispersions are so large that many faint
features in the spectrum are washed out, a considerable amount of information between 4200 and 6400 \AA\ is contained
in these indices. This library since then has been by far the most popular stellar library, and has been used many
times to derive stellar population parameters of galaxies. Since the library is quite noisy, special {\it fitting
functions} were made for every index, to facilitate the interpolation in the library when calculating SSP models (see
e.g. Worthey et al. 1994).

With the advent of linear CCDs soon more accurate stellar libraries became available, e.g. the library of Jones et
al. (1997), which contains spectra, not just indices, covering 2 small wavelength regions. This library was used by Vazdekis (1999), who created the first
stellar population models producing continuous output spectra. Now, 13 years later, among the most-used stellar libraries
are MILES (S\'anchez-Bl\'azquez et al. 2006) and Elodie (Prugniel et al. 2007), although many other libraries exist,
especially in the optical. The coverage of these libraries in terms of T$_{eff}$ and log $g$ is very good around solar
metallicities. Outside the solar metalliticy regime, the range of  stellar population ages that can be modeled is
somewhat restricted. The main problem at the moment is, that the wavelength range of these models is limited (e.g.
the wavelength coverage of MILES is 3500-7500\AA), and that only Galactic stars are included, most of them in the
Galactic disk, implying that their abundance ratios reflect the formation history of the Galactic disk, with
abundance ratios around solar. Recently, the MILES, CaT (Cenarro et al. 2003) and INDO-US libraries (Valdes et al.
2004) have been merged into the MIUSCAT library (Vazdekis et al. 2012), homogenizing resolution and flux calibrating,
extending the library to 9500\AA. The resolution of the current libraries goes from R~=~$\Delta\lambda/\lambda$~=~40000 (Elodie) to about 2.5\AA
(MILES). A higher resolution, but still unpublished library is the UVES-POS stellar library (see Bagnolo et al.
2003), which has a resolution of 80000, but is not flux calibrated.

In the near-UV and the NIR, spectral libraries are smaller, with smaller ranges
in stellar parameter space,  but the situation is rapidly improving. In the
near-UV the best library is (and will remain for some time) NGSL,  the Next
Generation Spectral Library of Gregg, Silva and collaborators, containing
spectra of 382 stars taken with HST/STIS at a wavelength-dependent resolution of
2000-10000, covering 1670--10250 \AA\ with excellent flux calibration. In the NIR, a stellar library of 230 stars,
the IRTF spectral library  has recently become available:  a compilation of
stellar spectra that cover the wavelength range from 0.8 to 2.5 $\mu$m at a resolution of $\sim$ 7\AA\ and for 
some from 0.8 to 5.0 $\mu$m (Rayner et al. 2009; Cushing et al. 2005). The
library, which has been flux calibrated using 2MASS photometry mainly consists
of late-type stars  (F-M), AGB, carbon and S stars, and L and T dwarfs. Most of
the stars are of solar-metallicity. Apart from this there is the stellar
spectra of cool giants (Lan\c{c}on \& Wood (2000), which contains $\sim100$
stars taken at a resolution of $R=1100$.

Theoretical stellar spectra are reasonably reliable for stars down to G-type,
since in later types molecular bands, which are still hard to model, are
starting to dominate the spectrum. Colors in the visible and near-IR bands are
reproduced within the error bars for temperatures down to  3500K. At a
resolution of R = 10$^5$ , the spectrum of the Sun is today reproduced to 5\% of
its relative flux, Arcturus (K1.5III) is reproduced to 9\% and Vega (A0V) is
reproduced to 1\%. Residuals are in general larger towards cooler stars or lower
surface gravities. For stars below  3500K, current developments in
hydrodynamical models and pulsating atmospheres should improve the accuracy of
the models, but it may well take some years before such grids are available to
the completeness needed for population synthesis. For O- and B-type stars,
recent developments in mass loss modeling, expanding atmospheres and wind
features are being incorporated into the theoretical grids to model the UV with
better accuracy (see Coelho 2009). Theoretical (or possibly semi-theoretical) libraries are the
most promising to model the integrated spectral features of galaxies. As our
understanding of stellar spectra improves,  more and more analysis will be done
with theoretical stellar libraries, especially in the optical and  the near-UV.
In the near-IR, I expect empirical libraries to remain highly superior for years
to come. Progress should come from libraries with at the same time high spectral
resolution and large wavelength coverage, such as the future X-Shooter spectral
library XSL (Chen et al. 2011). By analyzing galaxies using at the same time
optical, UV and  near-IR features one should be able to constrain the old
population dominating the mass, as well as  recent or intermediate bursts of
star formation at the same time.

\subsection{The Initial Mass Function}
\label{IMF}

An important ingredient in stellar population models is the Initial Mass Function. The IMF ultimately
determines the mass to light ratio of a stellar population, and therefore choosing the right IMF is
important to determine, among other things, the amount of dark matter in a galaxy. 

\begin{figure}
\begin{center}   
\includegraphics[height=8cm,clip]{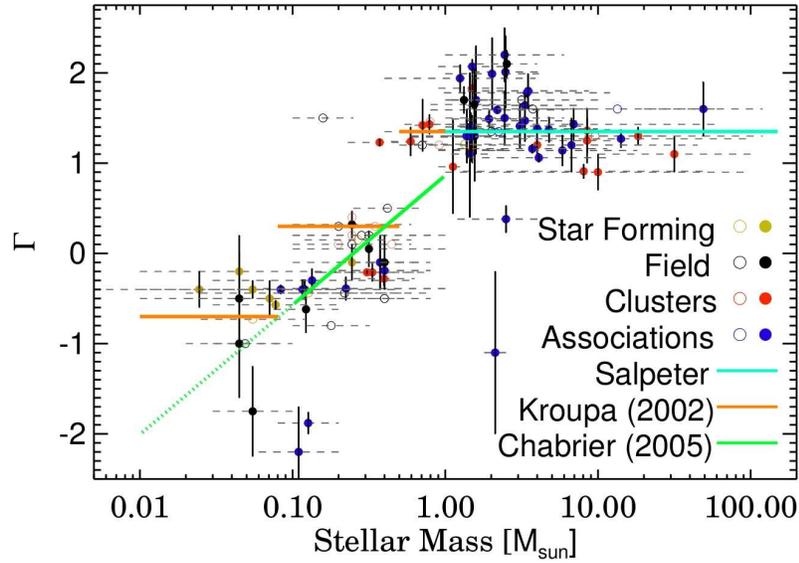}
 \caption{Derivative of the IMF-slope ($\Gamma$) as a function of stellar mass, derived from individual star counts (from Bastian et al. 2010). The colored solid lines represent three analytical IMFs: the Chabrier (2003) IMF (green), Salpeter (1955) in blue, and Kroupa (2001) IMF in orange.}
\label{IMFFig}
\end{center}
\end{figure}

Generally the IMF is determined by determining the {\it mass function} of stars and correcting them for
their evolution, to establish the distribution of stars of a stellar population at birth. This can be done
for stars in the solar neighborhood, or in Galactic clusters. A huge research effort has been invested to
determine the shape and variability of the IMF. Many details about this can be found in e.g. the review by
Kroupa (2007).   Salpeter (1955) first described the IMF as a power-law, $dN ~=~ c~m^{-\alpha}~dm$, where
$dN$ is the number of stars in the mass interval $m$, $m+dm$ and $c$ is the normalization constant. By modelling
the spatial distribution of the then observed stars with assumptions on the star-formation rate,
Galactic-disk structure and stellar evolution time-scales, Salpeter determined $\alpha$ to be 2.35 for 0.4
$<$ $m$ $<$ 10 $M_\odot$. Later on, many other studies have extended the mass ranges to higher and lower
mass stars. The current situation is that the stellar IMF {\it appears to be extra-ordinarily universal},
with an IMF which still has more or less the same slope down to about 0.5 M$_\odot$, but turning over
towards lower masses (the Universal IMF of Kroupa, his equation (20), see Fig.~\ref{IMFFig}). Without such a turnover, the amount
of mass in a stellar population can become very large, which often gets into conflict with dynamical
measurements of mass in galaxies, since those should be at least as large as the mass in stars and stellar
remnants. Kroupa also argues that the evidence for a top-heavy IMF is not strong in well-resolved starburst
clusters, and that dynamical evolution of the clusters needs to be modelled in detail to understand
possibly deviant observed IMFs. For the low-mass end of the IMF one has a reasonable idea of the IMF-slope
down to about 0.2 M$_\odot$ from star counts in the solar neighborhood. Also, here, the evidence for
variations in the IMF is weak, although there might be some speculative systematic variation with
metallicity in the sense that the more metal-poor and older populations may have flatter MFs as expected
from simple Jeans-mass arguments. Bastian et al. (2010) in a review for Annual Reviews in Astronomy and
Astrophysics confirms Kroupa's ideas: {\it We do not find overwhelming evidence for large systematic
variations in the IMF as a function of the initial conditions of star formation. We believe most reports of
non-standard/varying IMFs have other plausible explanations}.

In Section \ref{IMFMass} we will look at evidence for different types of IMFs based on integrated stellar
populations. This evidence should be considered in the light of the above mentioned detailed work in our
Galaxy.

\subsection{Abundance ratios}
\label{abunrat}

\begin{figure}
\begin{center}   
\includegraphics[width=\textwidth,clip]{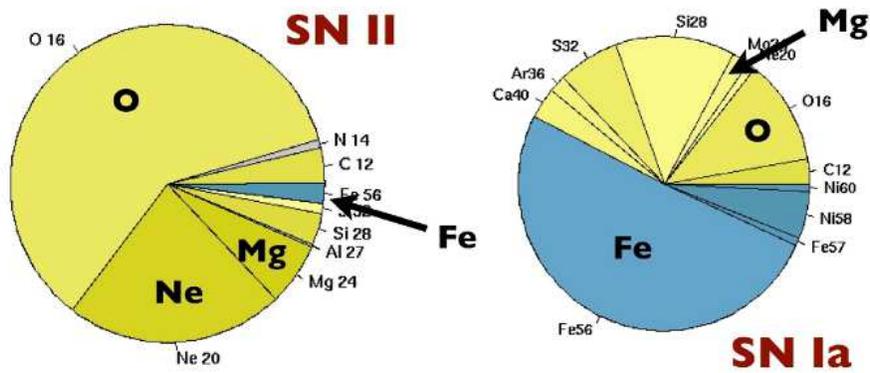}
\caption{Supernova yields (= relative mass of metals released into the ISM at
the death of the star), for SN type Ia and II. From Russell Smith, private communication.}
\label{snfig}
\end{center}
\end{figure}

Abundance ratios of elements in galaxies contain a wealth of information about the way the stars were formed, and their formation timescales. Element production for elements heavier than He is thought to take place in stars. Explosions of supernovae and stellar mass loss cause these newly formed atoms to enter the interstellar medium, where they can be used to form new stars, which in turn can again enrich the interstellar medium later. It is thought that Supernovae type II (massive stars) and Ia (C-degenerate white dwarfs) are responsible for most of the matter ejected into the ISM. SN type II eject mostly so-called $\alpha$-elements into the ISM (see Fig.~\ref{snfig}), i.e. O, Ne and Mg. SN type Ia, which take longer to produce, since the stars first have to evolve to their white dwarf phase, mainly produce Fe-peak elements. The fact that the timescale of element production in these supernovae is so different makes it possible to use the abundance ratio of an $\alpha$-element and a Fe-peak element as a star formation {\it clock}. A galaxy which forms all its stars early-on will have a much higher $\alpha$/Fe ratio than a galaxy that forms its stars slowly. 

\begin{figure}
\begin{center}   
\includegraphics[height=8cm,clip]{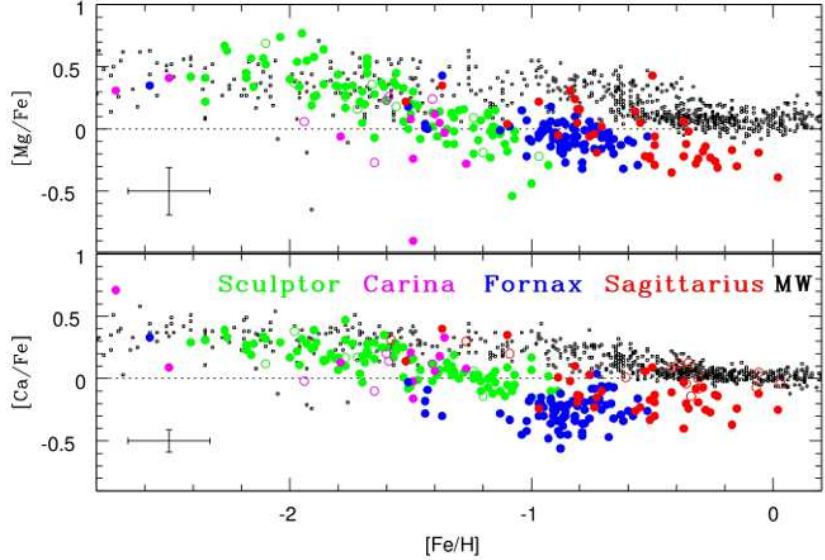}
 \caption{Abundance ratios of Mg/Fe and Ca/Fe for the Milky Way and a few local group dwarfs (from Tolstoy et al. 2009). For the origin of the individual measurements see Tolstoy et al. 2009. }
\label{abratlg}
\end{center}
\end{figure}

In our Galaxy it has been possible to measure abundances of many elements since almost 20 years (Edvardsson et al. 1993). This paper shows that the [Mg/Fe] ratio for disk stars is higher than solar for low metallicity stars, and at a certain metallicity starts to flatten off towards the solar value. This means that when our Galaxy was formed, and the metallicity of stars was low, they formed quickly. After a number of generations of metal enrichment, star formation became slower, and element production by SN type Ia started taking over. A nice review about abundances and abundance ratios in our Galaxy and in the Local Group is given by Tolstoy et al. (2009). In Fig.~\ref{abratlg} the current status of Mg/Fe measurements in the Local Group is shown. Dwarf galaxies each have their own {\it knee} in the [Mg/Fe] vs. [Fe/H] diagram, indicating the start of SN Ia element formation. The position of the knee indicates the metal-enrichment achieved by a system at the time SN Ia start to contribute to the chemical evolution (e.g., Arimoto \& Yoshii 1987, and modelled by e.g. Matteucci \& Brocato, 1990). This is between 10$^8$ and 10$^9$ yr
after the first star formation episode. A galaxy that efficiently produces and
retains metals over this time frame will reach a higher metallicity by the time
SN Ia start to contribute than a galaxy with a low star formation rate. 

Tolstoy et al. also explain that differences in relative abundances are to be expected between various $\alpha$-elements. Conditions related to SN II explosions are responsible for the fact that O and Mg often show different trends with [Fe/H] than Si, Ca and Ti (e.g., Fulbright, McWilliam \& Rich, 2007). 
Apart from elements ejected into the ISM by SNe, also giant and supergiant stars are responsible for the enrichment of the ISM. 
For these evolved stars (giants or super-giants) material synthesized in the core will have been mixed to the surface layers, and ejected into the ISM through e.g. dust. Most of the material ejected in this way is C, N and O.

The production of heavy neutron capture elements is less well understood. One distinguishes 2 processes, depending on the rate at which these
captures occur, and are called the slow (s-) and the rapid (r-) process. 
The s-process occurs in low to intermediate-mass 
thermally pulsating AGB stars (see Travaglio et al., 2004, and references
therein). R-process production is associated with
massive star nucleosynthesis. More about these elements and possibilities about using their abundances as star formation clocks can be found in Tolstoy et al. 

I will discuss abundance ratios from unresolved galaxies in Section \ref{spectralanalysis}.

\section{Stellar population analysis of individual galaxies}
\label{indgal}

Having described the ways that stellar population models are being made, with some caveats about their capabilities, I will give some examples of their use in this sectionFrole, indicating a number of issues that one has to be aware of when applying them. I will first discuss stellar population analysis on colors, and then on line indices or continuous spectra.

\subsection{The age-metallicity degeneracy and luminosity weighting}
\label{agemetdeg}

\begin{figure}
\begin{center}   
\includegraphics[height=8cm,clip]{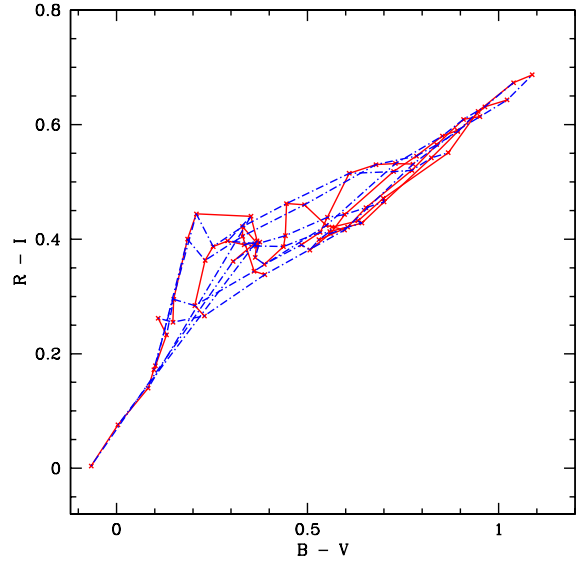}
 \caption{The age-metallicity degeneracy as seen in a color-color diagram. Shown
 is a grid of SSPs with varying metallicity and age. Especially in the right part of the diagram age and metallicity cannot be determined independently for given observations of $B-V$ and $R-I$. Used here are the MILES models (Vazdekis et al. 2010) with unimodal IMF and slope $x$=1.3.}
\label{agemet} 
\end{center}
\end{figure}

Determining both an age and a metallicity of a galaxy, or even of an SSP, is
tougher than it seems.  Galaxy colors become redder as the galaxy ages, since
more stars move to the giant branch, and  also for increasing metallicities,
since the effective temperatures of most stars decrease because of increasing
opacities in the stellar photosphere. Colors and many line strengths in the
optical basically  depend on the temperature of the main sequence turnoff. The
effects of increasing the age can be compensated  for many observables by
decreasing the metallicity. Worthey (1994) estimated that a factor of 3 increase
in metallicity corresponds to a factor of 2 in age when using optical colors as
age indicators, the so-called 2/3 rule. Optical colours are notoriously degenerated (see Fig.\ref{agemet} in the red part of the diagram).

There are, however, ways to break the degeneracy. Younger stellar populations
with higher metallicities have a bluer contribution from the main sequence
turnoff, and a redder one from the RGB. By using an optical color, together
with a color that has a much higher relative sensitivity to the turnoff stars
the degeneracy can be broken. This would be the case using colors such as UV --
V, or Balmer line indices such as H$\beta$ or H$\gamma$. Equivalently, a
combination of an optical color or line index which is strongly sensitive to
the contributions of very cool giants would also break the degeneracy. Such
colors would be e.g. V -- K or J -- K, or the CO index at 2.3 $\mu$m. 

\begin{figure}
\begin{center}   
\includegraphics[width=\textwidth,clip]{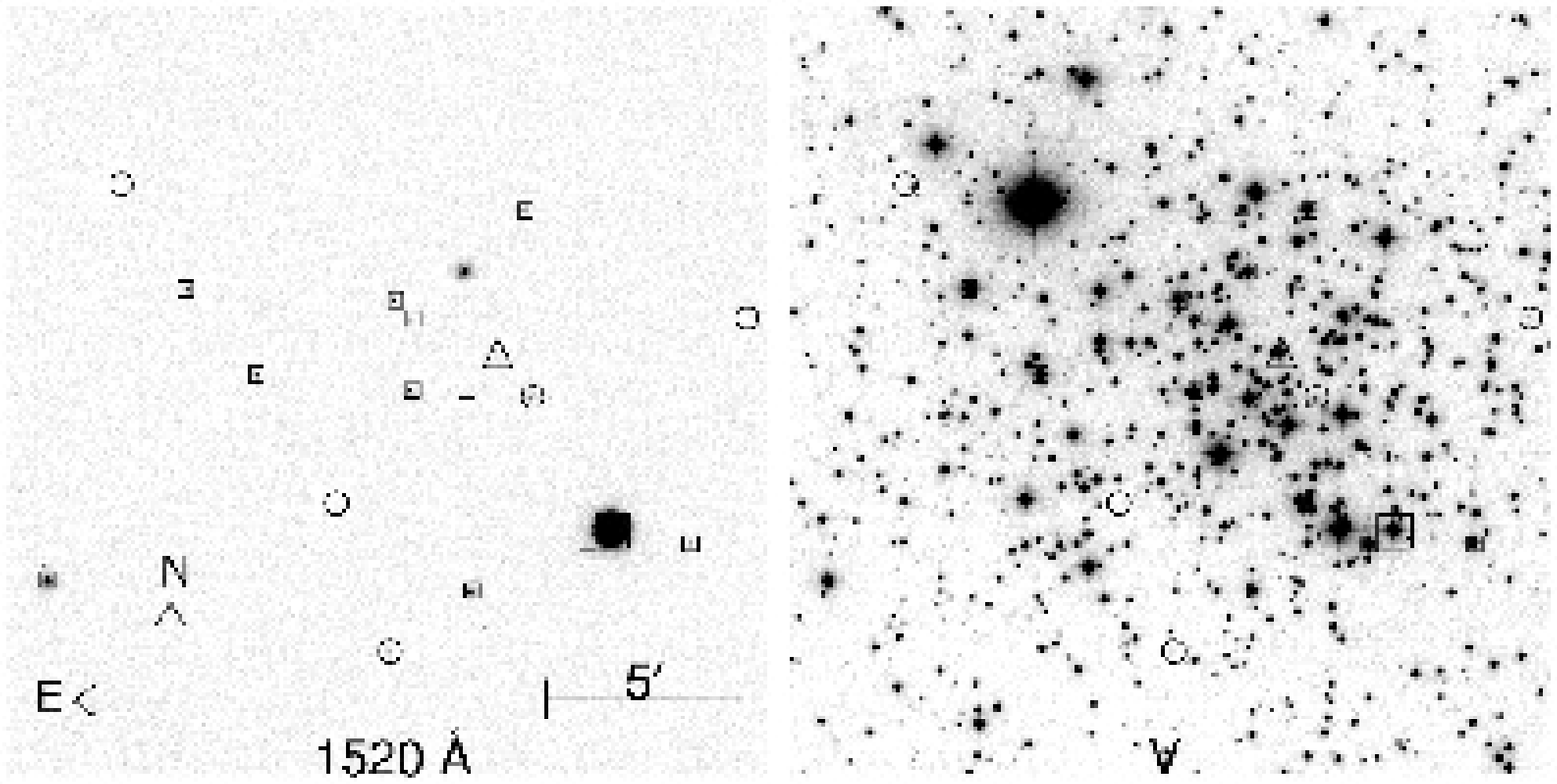}
 \caption{UIT image at $\sim$ 1500\AA\ and in $V$ of the open cluster M67 (Landsman et al. 1998). Note that one star completely dominates the light in the UV. }
\label{uvlight}
\end{center}
\end{figure}

Both methods are being applied. For spectra covering only a small range in wavelength, very sophisticated
indices have been developed maximizing age-sensitivity while minimizing the sensitivity to metallicity
(e.g. Vazdekis \& Arimoto 1999). In general, assuming that the stars in a galaxy are not coeval, a blue
spectrum will give a different mean age than a red spectrum, since colors/indices in the blue will be more
sensitive to the younger stars etc. This is the so-called {\it luminosity weighting} of stellar populations
(I prefer not to use the term {\it light weighting}, since this has other associations, in e.g. material
sciences). When applying stellar population synthesis codes, one should always realize that one's results
have been weighted with the luminosity of the stars, implying that the brightest stars give the impression
to be more important than they really are, when one weights according to mass, the natural choice. As an
example, in the UV, sometimes 90\% of the light in a cluster is coming from one star (Landsman et al. 1998)
(see Fig.~\ref{uvlight}). This means that the mass-weighted age of that cluster could in principle be very
old, while the luminosity weighted value is close the value for that star, i.e. young. For the 
interpretation of galaxy ages this distinction between mass and luminosity weighting is particularly
important. 

\subsection{Analysis using colors, and the role of extinction}
\label{colouranalysis}

\begin{figure}
\begin{center}   
\includegraphics[width=\textwidth,clip]{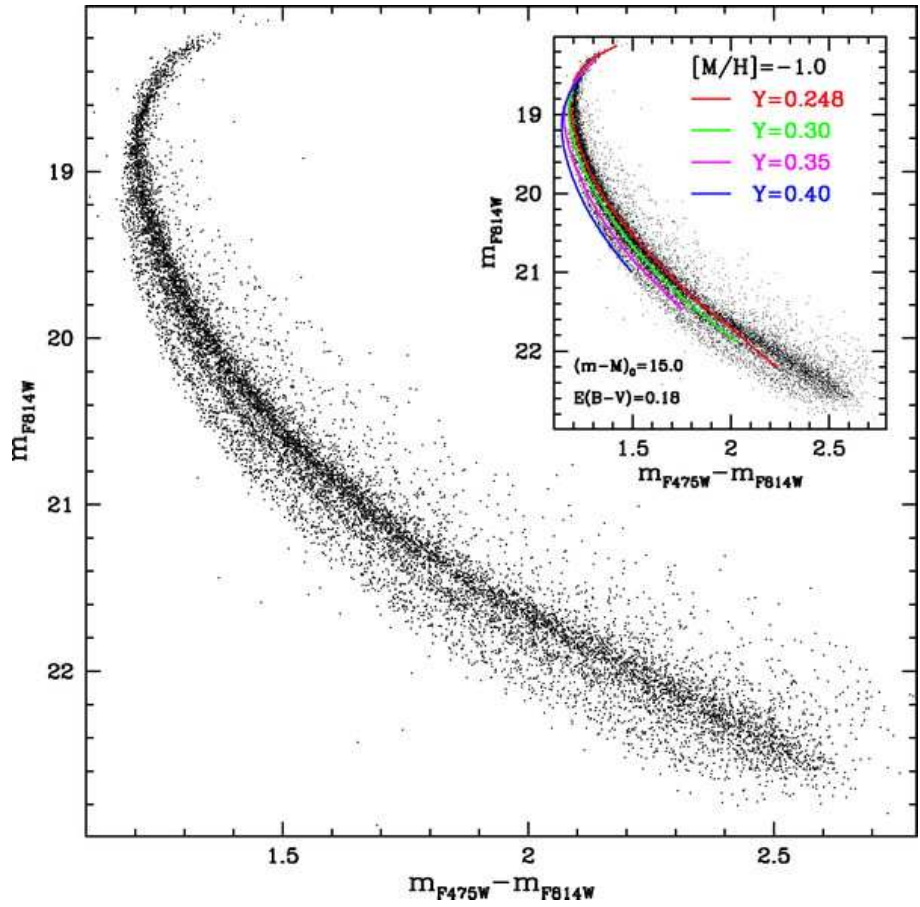}
 \caption{Differential reddening corrected CMD of the globular cluster NGC 2808. In the inset some fits have been done with models of age 12.5 Gyr and various He abundances (from Piotto et al. 2007).}
\label{ZAMS}
\end{center}
\end{figure}

In Chapter \ref{models} we have found out that stellar populations in globular clusters are SSPs, and that
populations in galaxies can be considered as linear combinations of SSPs. Recently, however, we have
learned that the first assumption does not always hold. For a while it has been known that  $\omega$ Cen,
which up to now was considered to be a globular cluster, shows a spread in metallicity and possibly also age
(Norris \& Da Costa 1995). Conservative people could maintain for another 10 years that globular clusters
have a single metallicity, by claiming that $\omega$ Cen is a galaxy, until recently Piotto et al. (2007), see Fig.\ref{ZAMS},
discovered multiple main sequences in the globular cluster NGC~2808. At the same time Mackey \& Broby
Nielsen found multiple main sequences in the LMC cluster NGC~1846. More clusters have been found later
showing similar effects (e.g. Milone et al. 2008, Mackey et al. 2008). It is not clear yet what the reason
is of these multiple branches. It could be that the He (or CNO) abundance is different, but also there might be a
difference in age/metallicity. Spectroscopic studies here will have to show what really is happening.

\begin{figure}
\begin{center}   
\includegraphics[width=\textwidth,clip]{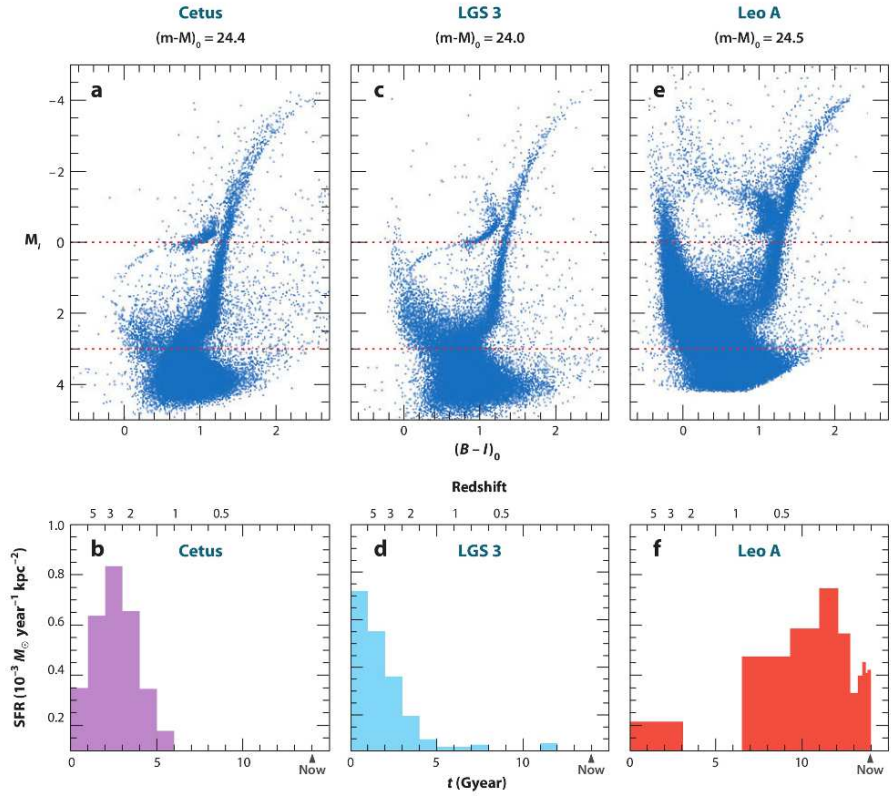}
 \caption{HST/ACS color-magnitude diagrams SFHs for three Local Group dwarf galaxies: Cetus, a distant dwarf spheroidal galaxy LGS 3, a transition-type dwarf galaxy  and  Leo A, a dwarf
irregular.  These results come from the LCID project (Gallart \& the LCID team 2007, Cole et al. 2007). From Tolstoy et al. (2009).}
\label{sfhist}
\end{center}
\end{figure}

Color-magnitude diagrams can not only be used for globular clusters but also  for galaxies in the rest of
the Local Group. With HST it is possible to resolve individual stars below the Main Sequence Turnoff, and
this way obtain exquisite star formation histories. In Fig.~\ref{sfhist} I have reproduced a figure from
the review of Tolstoy et al. (2009) with star formation histories of 3 dwarfs. An earlier, also excellent
review, is by Mateo (1998). The star formation histories show that there are large variations between the
galaxies of the local group, even between galaxies that have the same morphological classification (M32,
NGC~205 and NGC~185). There are no galaxies for which we can exclude the presence of an underlying old
population. Radial gradients in the populations of individual galaxies are seen as well. As mentioned
before, more information about the abundance ratios in individual stars, giving information about star
formation timescales, can be obtained from spectroscopy of bright giants in these galaxies. 

When one goes further away, one can only resolve stars on the Red Giant Branch
and beyond. One can obtain the spread in metallicity, e.g. in Centaurus A
(Harris et al. 1999), and the galaxies in the ANGST survey (Dalcanton et al.
2009). A great application of counting the stars on the RGB is to use these star
counts to make maps of the stellar density in the outer parts of galaxies. This
way people have found huge low surface brightness features linking M31 with its
companions, including M33, probably remains from encounters between these
galaxies (Ibata et al. 2001, McConnachie et al. 2009). For the spiral galaxy
NGC~300 Bland-Hawthorn et al. (2005) have been able to measure the stellar
surface brightness profile to a distance of 10 effective radii from the galaxy
center in this way.

\begin{figure}
\begin{center}   
\includegraphics[width=\textwidth,clip]{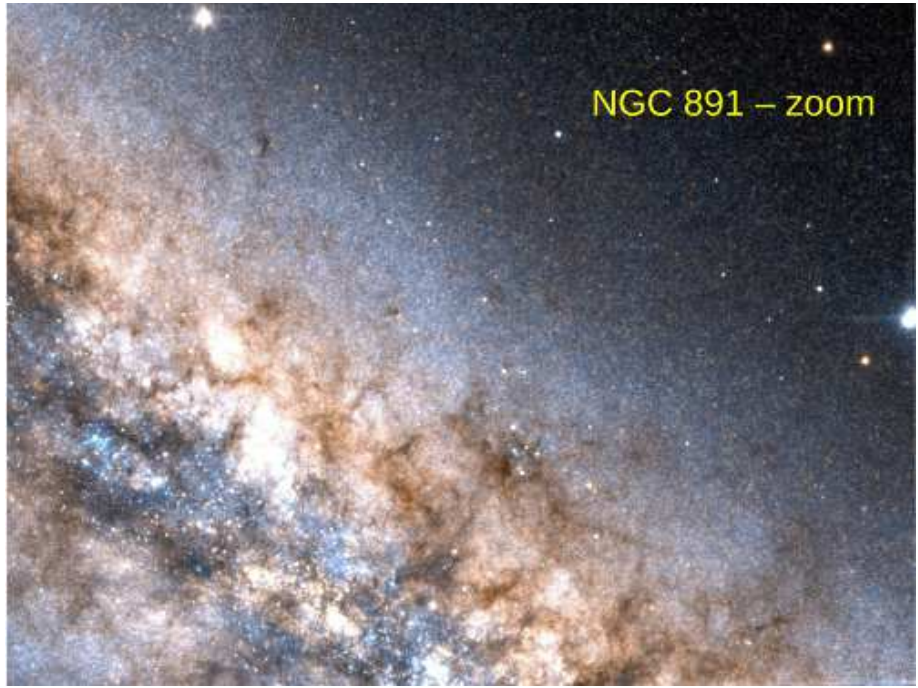}
 \caption{ Composite HST F814W/F555W image of NGC~891 (from the Hubble Legacy Archive). This galaxy is sometimes called the {\it twin} of our Milky Way. Note the young stars in the mid-plane and the dust extinction filaments.}
\label{N891}
\end{center}
\end{figure}

\begin{figure}
\begin{center}   
\includegraphics[height=8cm,clip]{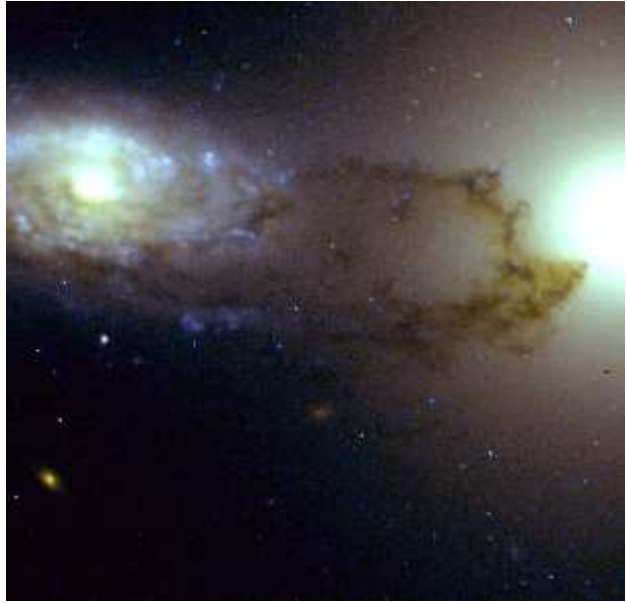}
 \caption{Dust in a spiral galaxy seen in absorption in front of an elliptical galaxy (from White \& Keel 1992).}
\label{whitekeel}
\end{center}
\end{figure}

In Fig.~\ref{N891} a closeup is given of an RGB image of the disk of NGC 891, a
nearby edge-on galaxy. What is clear are the many bright stars in the disk of
the galaxy. Above it, many red filaments are seen. They are dust-lanes, seen up
to large distances from the plane. In the lectures by Daniela Calzetti (this
volume)  you can see a lot of material about this dust, and how it extincts the
light behind it. In Fig.~\ref{N891} for example, further study shows that the
blue stars seen in the left bottom corner are found in front of most of the disk
of the galaxy, which itself is barely seen because of the extinction. In
Fig.~\ref{whitekeel} one can see that  the extinction is usually associated to
spiral arms, and that it can be present to large radii. Here the extinction in a
spiral disk is seen in front of an elliptical galaxy.

Dust extinction is found predominantly in spiral galaxies of type Sab-Sc (e.g.
Giovanelli et al. 1994). It is generally associated to molecular gas, and is
stronger in larger (higher metallicity) galaxies. The UV energy absorbed by the
dust is re-radiated in the IR and submm, responsible for a large fraction of the
emission at those wavelengths. As far as stellar population synthesis is
concerned the most important effect is that it reddens the colors using the
dust extinction law (e.g. Cardelli et al. 1989). Reddening is strongest in the
blue, and almost non-existent red-ward of 2 $\mu$m. In our Galaxy, it is
impossible to see the Galactic Center in the optical, because of more than 20
magnitudes of extinction. However, in the infrared, at 2 micron, the extinction
is only about 2 mag, so that observations there are easily possible. The ratio
of reddening of dust in various colors is very similar to the effect of
metallicity (and even age). This means that by simply measuring two colors, one
cannot correct for the effects of extinction. For that more colors, or a
spectrum are needed. It is therefore also not easy to measure the extinction
from colors in a spiral galaxy. If one wants to do this, one can e.g. measure the amount of
extinction statistically, by looking at the dependence on inclination. The
color of a galaxy without extinction should not depend on inclination, while
for an inclined galaxy the path length through the dust is longer, and thus the
extinction. In Peletier et al. (1995) this technique is used to show that many
nearby spiral galaxies in the $B$-band are optically thick within their central
effective radius, but not in their outer parts. Certainly in bulges
extinction plays a large role in many galaxies, and people analyzing the colors
of external bulges should take this into account (e.g. Balcells \& Peletier
1994).

\begin{figure}
\begin{center}   
\includegraphics[width=\textwidth,clip]{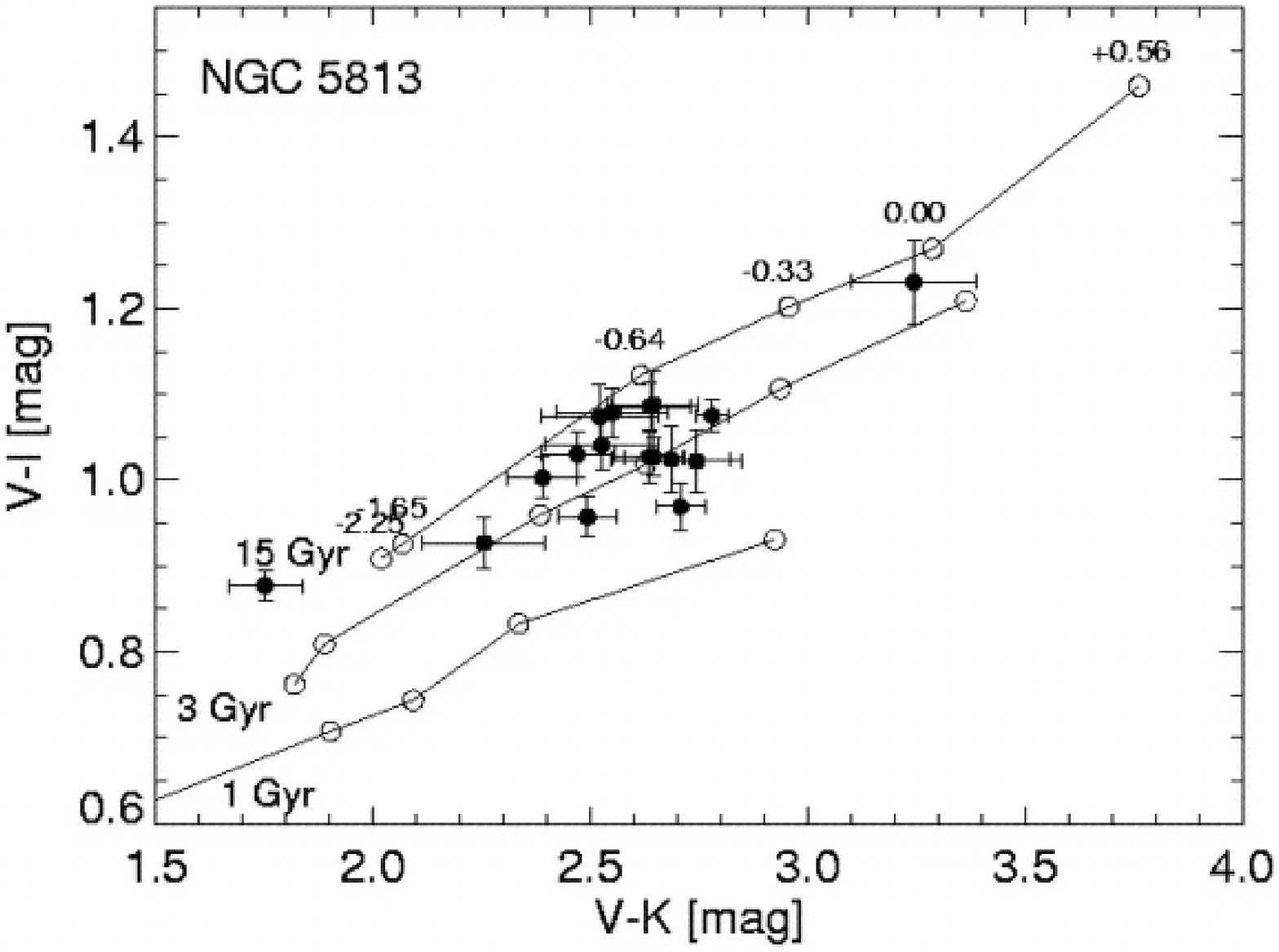}
\includegraphics[width=\textwidth,clip]{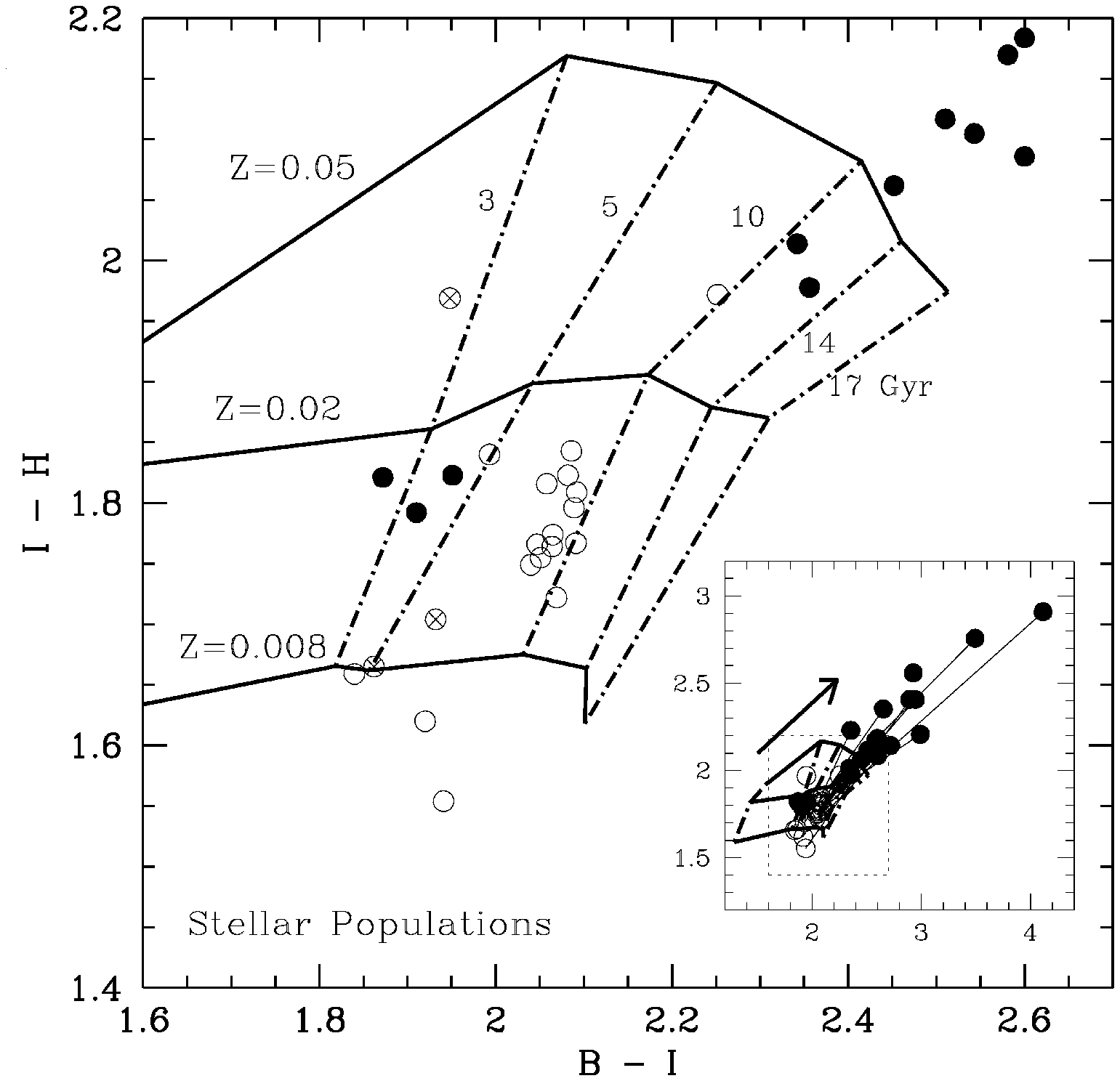}
 \caption{Top: $V-I$ vs. $V-K$ diagram for globular clusters in NGC~5813 (from Hempel et al. 2007). Bottom: $B-I$ vs. $I-H$ diagram for bulges of early-type spirals  (from Peletier et al. 1999).}
\label{colcol}
\end{center}
\end{figure}

In the absence of dust color-color diagrams consisting of an optical and an
optical-IR (or IR) color, or, e.g., of a UV-Opt and an optical color, can be
used to separate the effects of age and metallicity. This method is popular for
globular clusters in nearby galaxies, for which high quality spectra are
difficult to get. As can be seen in Fig.~\ref{colcol}, it is important that
accurate observational colors are available. Such studies can maybe explain the
bi-modality in globular cluster colors (Ashman \& Zepf 1992). Chies-Santos et
al. (2012), for a sample of 14 early-type galaxies, found that, although the
optical color distributions of the globular clusters are bimodal, this is not
always the case in the infrared (z-K). The authors explain this results with a
non-linear color-metallicity conversion, but clearly state that better data are
required to confirm their results. This means that colors of globular clusters
are not well understood. The, up to recently, firm conclusion that globular
clusters have a bimodal metallicity distribution is now up for discussion.

The same color combination is also often used for galaxy research. In
Fig.~\ref{colcol} (right) HST data are shown of the inner parts of a sample of
early-type spiral galaxies (Peletier et al. 1999), on top of a grid of SSP
models. Central colors are indicated with red, filled symbols, while the
colors at 0.5 bulge effective radii are indicated in open, blue circles. These
latter colors are calculated on the minor axis of the bulge, on the side not
obscured by the galaxy disk. Interesting to see from this plot is that most
galaxy centers lie, often far, away from the model grid, indicating considerable
amounts of extinction A$_V$ of often more than 1 magnitude. The blue points
cluster mostly together. Although the exact position of the model grid should be
taken with caution (color differences are probably more reliable than exact
colors, because of systematic effects in the models), this diagram seems to
indicate that the bulges in this sample are old (mostly around 8-9 Gyr) with
metallicities somewhat below solar.  

The fact that one has to determine such detailed colors leads to another
problem: the models. Up to recently, the spectrophotometric quality of the
stellar population models has not been very high. In S\'anchez-Bl\'azquez et al.
(2006) it is shown that the colors of stars in the MILES sample are consistent
with their spectra within 1.5\%, something which cannot be said from e.g. the
Stelib library (Le Borgne et al. 2003), used for the Bruzual \& Charlot (2003)
models. Vazdekis and his group have extended the MILES library with a subset of
good spectra from the Indo-US library (Valdes et al. 2004), with the aim of
making a stellar library with wavelengths up to 9500\AA\, with good flux
calibration. In Ricciardelli et al. (2012) they show that, when fitting
well-calibrated SDSS data, there are problems fitting the $g-r$ vs. $r-i$
distribution of galaxies for galaxies with high velocity dispersions. In this
paper it is discussed that probably 
$\alpha$-enhanced stellar population models are needed to solve this issue.  

\subsection{Analysis using optical spectra}
\label{spectralanalysis}

The main difference between a spectrum of a star and one of a composite system,
such as a galaxy, is that the composite spectrum is the sum of many stellar
spectra, weighted by their flux at the particular wavelength, and shifted by
their individual radial velocities. These velocity shifts are not to be
discarded. In a large galaxy stars move through one another with a velocity
dispersion of about 300 km/s. This means that every line in the spectrum is
broadened by a Gaussian with such a dispersion, which means that abundances can
not be measured any more from narrow lines from single transitions, since those
are all blended. Abundances have to be obtained by fitting stellar population
models with given abundances to the galaxy spectra. Velocity broadening cannot
be avoided, and we have to live with broadened lines. The most common way to
measure line strengths in composite spectra is by using a system of line
indices. These indices are defined by three pass-bands: a feature band, and two
continuum bands, and measured as equivalent width: the surface (in \AA\ ) under
the spectrum that is normalized using the continuum on both sides (see Worthey
et al. 1994 for definitions of the Lick/IDS system. In the Lick/IDS system 21
indices were defined to measure the strongest stellar features in the spectrum
in the optical at a resolution of about 9\AA\ (Worthey et al. 1994). 4 more
indices (2 H$\gamma$ and 2 H$\delta$ indices) were added in 1997 by Worthey \&
Ottaviani. These indices were used by Trager et al. (2000) to determine stellar
population parameters from Lick/IDS indices of many nearby galaxies. Later-on, many more indices were added by Serven et al. (2005). Other
indices are available in the literature. Rose (1994) defined several indices
with a one-sided continuum, mainly in the blue. Cenarro et al. (2001) defined
indices in the region of the Ca II IR triplet, sometimes with with multiple
continuum regions. Normally an index becomes larger as the absorption line
becomes stronger. Some lines, however, like the H$\delta$ and H$\gamma$ lines in
the Lick system, are situated in such crowded regions, that their continuum
fluxes are affected by metal abundances, and that the line index sometimes can
be negative, even though H$\gamma$ or H$\delta$ is found in absorption.

Lick indices are hard to measure. Not only do they require the observed (galaxy)
spectrum to be convolved to exactly the right resolution, and do they require a
rather uncertain correction for velocity broadening, they also need certain zero
point corrections to make sure that the instrumental response of the observed spectrum has
the same shape as the Lick/IDS in the 1980's, when the standard stars for the
Lick system were observed. This is a tedious job, since the Lick system does not
work with flux calibrated spectra, and requires that for every observational
setup a number of Lick standards are observed. To improve the situation, a
slightly modified line index system (LIS) has been defined by Vazdekis et al.
(2010). It is based on the MILES stellar library, uses the same wavelength
definitions as the Lick/IDS system, and is defined on flux calibrated spectra,
so that it is much easier for people to use this backward compatible system. To
make it possible to also use less broad indices, for e.g. globular clusters and
dwarf galaxies, the LIS system has been defined for standard resolutions of 5,
8.4 and 14 \AA\ FWHM.

\begin{figure}
\begin{center}   
\includegraphics[height=8cm,clip]{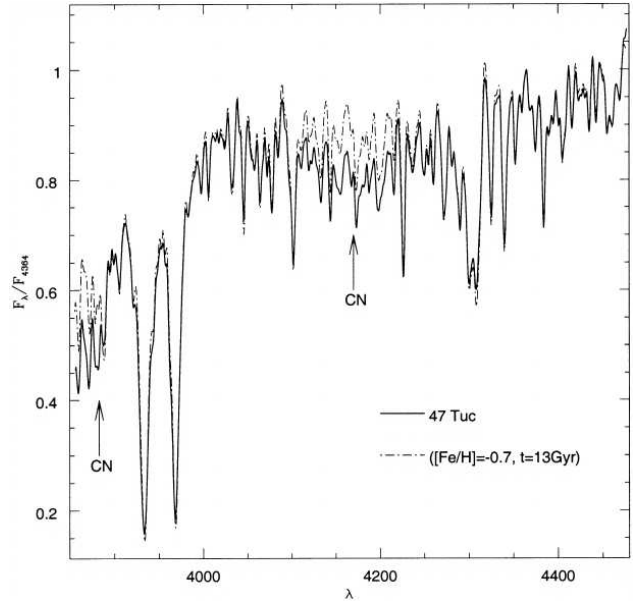}
 \caption{Blue spectrum of 47 Tuc, together  with an SSP fit by Vazdekis (1999). Marked are the well-known CN strong bands of this globular cluster.}
 \label{specfit}
\end{center}
\end{figure}

Although line indices are a good way to measure the strength of certain spectral
features, there is, at present, no need any more to go through the indices, when
comparing galaxy data with models, since one can directly fit the models to the
data. Vazdekis (1999) already showed the power of this method, which
dramatically can show regions of the spectrum where the models are inadequate
(see Fig.\ref{specfit}). Of course, the stellar population models will have to
be convolved with the correct LOSVD (Line of Sight Velocity Distribution), i.e.
the broadening of the stars. The SAURON team (Sarzi et al. 2006) have applied
full spectral fitting to Integral Field Spectroscopy of ellipticals and S0s,
fitting the observed spectrum at every position in every galaxy with a set of
SSP models, determining the LOSVD, and a best-fit stellar population model. They
noticed that always  large residuals occurred at the position of emission lines
such as H$\beta$ and the [OIII] line at 5007\AA. Realizing that they could at
the same time also measure the velocity and broadening of the ionized gas, they
then developed a method to fit at the same time velocity broadened SSP models
and Gaussians representing the emission lines to the data. As a result, ionized
gas could be found in 75\% of the sample, much more than was previously known.
This method is much more sensitive than e.g. methods that map emission lines
using narrow-band filters. Of course, those latter methods can cover a larger area, at
generally a higher spatial resolution. 

Separating absorption and emission is in particular important for lines which
are at the same time important absorption and emission lines, such as the Balmer
lines H$\beta$ and H$\gamma$. These lines, which are so important to determine
ages of stellar populations, can easily be filled in by emission. Before full
spectral fitting was possible assumptions were made that the H$\beta$ emission
line strength was a constant fraction of the [OIII] 5007\AA\ line. The maps of
Sarzi et al. (2006, 2010) show, however, that this ratio varies from galaxy to
galaxy. Methods like this are so powerful that Balmer absorption line strengths
can be measured in spiral galaxies with strong emission lines (e.g.
Falc\'on-Barroso et al. 2006, MacArthur et al. 2009).   

\begin{figure}
\begin{center}   
\includegraphics[width=\textwidth,clip]{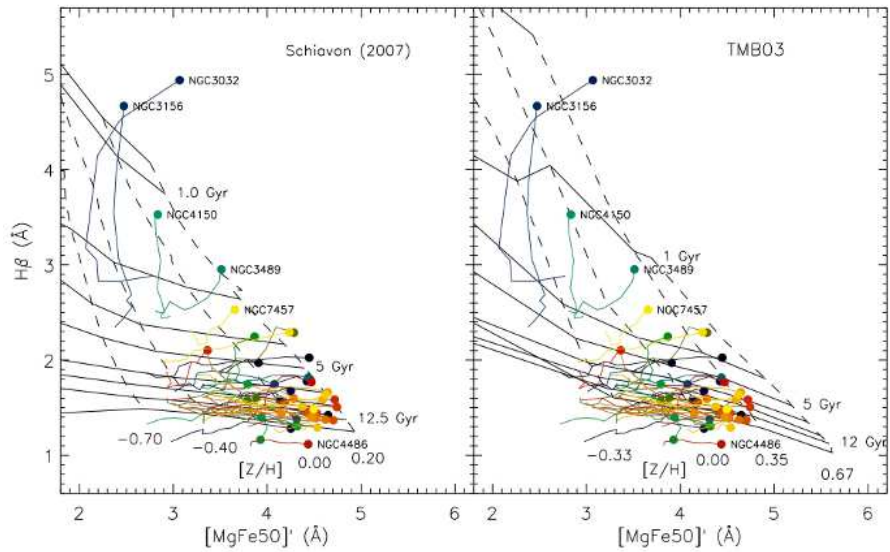}
 \caption{Radial line strength gradients for the 48 early-type galaxies in the SAURON sample. The center of each galaxy is indicated
by a filled circle and different colors are used for each galaxy. Over plotted are stellar population models by Schiavon (2007, left) and Thomas et al. (2003, right) for solar abundance ratios. Note
that the latter models extend to [Z/H] = +0.67, whereas Schiavon models reach only [Z/H] = +0.2 (from Kuntschner et al. 2010).}
\label{mgfehb}
\end{center}
\end{figure}

The most popular index-index diagram is the Mg$~b$ vs. H$\beta$ diagram. Here
Mg$~b$ is mostly sensitive to metallicity, while H$\beta$ is mostly
age-sensitive. Using this diagram it was found that massive galaxies have
$\alpha$/Fe ratios higher than solar (Peletier 1989, Worthey et al. 1992). In
Fig.~\ref{mgfehb} we show H$\beta$ against [MgFe50], a composite of the Lick
indices Mg~b and Fe 5015, from Kuntschner (2010). Here one sees how some
relatively small differences between stellar population models can change the
ages derived from these indices. Here the models of Schiavon (2007) and those of
Thomas et al. (2003) are shown. The galaxies are from the SAURON sample of
early-type galaxies, and are shown as lines, with a dot in the center. Using the
models of Schiavon, the old galaxies (these are mostly the ellipticals and the
massive S0's) are old everywhere, with metallicity decreasing outwards. If one
uses the models of Thomas et al. (2003), the outer parts are older, with similar metallicity
gradients. Note here that these ages are luminosity weighted ages, and that the
younger galaxies probably consist of an older continuum with a young population
superimposed. 

Measuring element abundance ratios from integrated spectra is much less
straightforward than for individual stars. What we know is that [Mg/Fe] varies
strongly as a function of galaxy velocity dispersion $\sigma$ (a proxy of mass).
We also know that $\alpha$-elements vary more as a function of $\sigma$ than Fe
(S\'anchez-Bl\'azquez et al. 2006). From a sample of galaxies in the Coma
cluster Smith et al. (2009) and Graves \& Schiavon (2008) claim that the abundance
ratios Mg/Fe and Ca/Fe simultaneously decrease with Fe/H and increase with
$\sigma$. These dependences can be explained by varying star formation time
scales as a function of $\sigma$, and therefore different ratios of element
enrichment by SN type II and Ia.  For both C/Fe and N/Fe, no correlation with
Fe/H is observed at fixed $\sigma$. This can be explained if  these elements are
produced primarily by low/intermediate mass stars, and hence on a similar
time-scale to the Fe enrichment. The element abundance ratio trends with [Fe/H]
are very similar to those in our Galaxy, which suggests a high degree of
regularity in the chemical enrichment history of galaxies (Smith et al. 2009).

\subsection{Star Formation Histories and the SSP}
\label{SFH}

The availability of full spectra makes it possible to recover the Star Formation
History (SFH) in some detail. Here one can divide the efforts into 2 parts:
efforts that fit the full spectrum, including UV and IR, using the far IR,
originating mostly from dust emission, as well as the submm on one hand, and
more detailed studies that determine the distribution of stars of various ages
on the other.  The most important result from the first kind of studies is the
amount of hot, young stars, responsible for ionizing the gas end heating up the
dust around them. For this work I refer to Daniela Calzetti's chapter in this
book. 

As far as the analysis of photospheric light is concerned, there is a growing
body of full spectrum fitting algorithms that are being developed for
constraining and recovering the Star Formation History (e.g., MOPED - Panter et
al. 2003; Starlight - Cid Fernandes et al. 2005; Steckmap - Ocvirk et al. 2006;
Koleva et al. 2008). SED fitting works best for large wavelength ranges. It is
based on a set of SSP models and an extinction curve, and fits at the same time
the stellar population mix, the LOSVD and the amount of extinction.

Using full spectrum fitting several independent bursts of star formation can be
determined. The number of parameters that can be recovered from a spectrum
depends strongly on the signal-to-noise ratio, wavelength coverage and presence
or absence of a young population (Tojeiro et al. 2007). However, the results are
strongly affected by the age-metallicity degeneracy, so interpreting the results
is difficult. Also, there is a certain degeneracy between the number of
components in the LOSVD, and the SFH. Some tests are shown in Ocvirk et al.
(2008), who try to constrain at the same time the SFH and the LOSVD along the
slit of the spiral galaxy NGC~4030. It is clear from this experiment that higher
spectral resolution and S/N data is needed to obtain astrophysically reliable
results which are not degenerate. 

On the other hand, such studies are very useful to determine whether a galaxy is
well fit by a one-SSP model, and if this is not the case, what the relative mass
fraction in the various components in the SFH is. That information is useful to
understand the formation history, combining it with spatial information about
e.g. interactions. I will give 2 examples here. 

\begin{figure}
\begin{center}   
\includegraphics[width=\textwidth,clip]{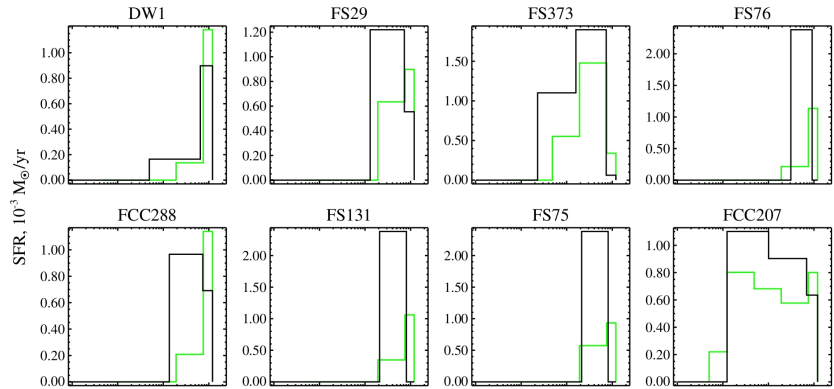}
 \caption{Star formation histories for the inner arcsec of a number of dwarf ellipticals, from Koleva et al. (2009). Black lines indicate SFR obtained using Steckmap, green using ULYSS.}
\label{koleva2009} 
\end{center}
\end{figure}

The first is by Koleva et al. (2009). They derive star formation histories of
dwarf ellipticals in the Fornax cluster using ESO-VLT data. To understand the
formation of these galaxies, it is very important to know whether these star
formation histories are extended or short-lived, and whether they are very
diverse, as is the case in the Local Group. In the Fornax cluster the
environment is different, with a stronger influence from the IGM. Usually the
photometric images are featureless, so that little can be learned about the
stellar populations from the morphological structure. The results show that the
star formation histories are not very different from those in the Local Group,
and vary from SSP-like to extended (Fig.\ref{koleva2009}). 

A second example is from MacArthur et al. (2009), who determined star formation histories in a sample of late-type spirals using long-slit high S/N Gemini/GMOS data. For these spirals, imaging already shows that they have composite stellar populations. This is confirmed and quantified by stellar population synthesis. The authors are able to derive star formation histories consisting of a number of logarithmically spaced SSPs. One of their most important conclusions is that, although young populations contribute a large fraction to the galaxy light of late-type bulges, in mass they are predominantly
composed of old and metal rich SPs (at least a mass fraction of 80\% ).

\subsection{Learning about stellar populations using 2D spectroscopy} 
\label{specIFU}

In the previous sections we have discussed how to derive the star formation history of a stellar population at a given position in a galaxy. One should always remember that galaxies are morphological and dynamical entities, and that the stellar populations that one derives are the result of the formation and evolution of the galaxy, and therefore intimately related to the morphological/kinematical component that one studies. 2-dimensional spectroscopy is an ideal tool to connect stellar populations with morphology and dynamics. In the last decade many of the large galaxies in the nearby Universe have been studied using the SAURON instrument at La Palma. The NIR Integral Field Spectrograph Sinfoni at ESO's VLT is making a large impact in the field at galaxy formation at z $\sim$ 2 (SINS survey - F\"orster-Schreiber et al. 2009). Many IFU surveys are being planned (e.g. CALIFA, S\'anchez et al. 2012; VIRUS, etc.). 

The SAURON survey (de Zeeuw et al. 2002, Bacon et al. 2001) has shown that many early-type galaxies contain kinematically-defined central disks. H$\beta$ absorption maps often show that these disks contain stars that are younger than the stars in the main body of the galaxy. The connnection here between the stellar populations, the morphology and the kinematics shows us that these disks are formed later. Since most of the disks have angular momentum that has the same sign as that of the main galaxy, one thinks that the disks are formed from gas lost by stars in the galaxies themselves (Sarzi et al. 2006). 
Spiral galaxies, much more than elliptical galaxies, have several components, such as bulge, inner disk, outer disk, bars, rings, etc. Here studying the stellar populations together with morphology and dynamics much more enhances our understanding of all these components, and the galaxies as a whole.   
I will discuss here the inner regions of 2 spiral galaxies, in order of morphological type, from the SAURON study of Falc\'on-Barroso et al. (2006) and Peletier et al. (2007).

\begin{figure}
\begin{center}   
\includegraphics[width=\textwidth,clip]{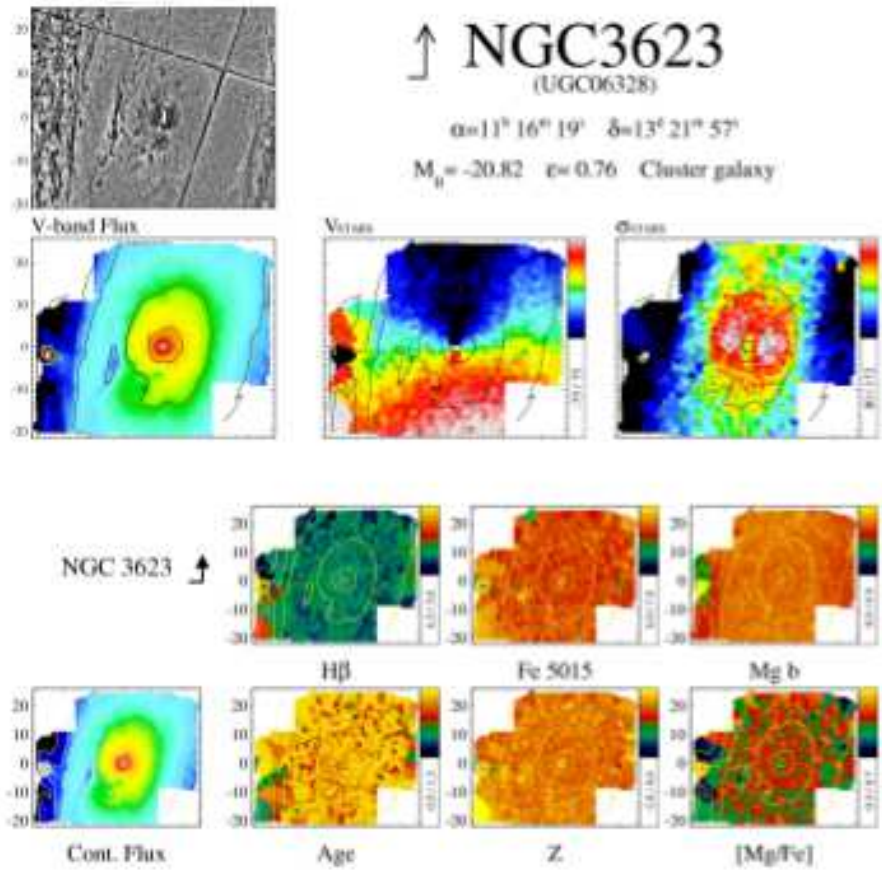}
 \caption{2-dimensional maps of various quantities in the inner regions of NGC 3623: (from top left to bottom right): unsharp masked HST image, V-band continuum image, stellar velocity field and velocity dispersion maps, H$\beta$, Fe~5015 and Mg~b absorption line maps, again V-band continuum image.  age, metallicity and [Mg/Fe] map. All from SAURON (Falc\'on-Barroso et al. 2006, Peletier et al. 2007), except for the HST image. }
\label{N3623}
\end{center}
\end{figure}

The first one is the Sa galaxy NGC~3623 (Fig.~\ref{N3623}). The inner regions of this galaxy mainly contain old stellar populations, as shown by the absorption line maps and confirmed by the unsharp masked HST image, an image which is a very good indicator of dust extinction. Since young stars are always accompanied by extinction, unsharp masked images are an efficient way to find younger stars. However, the presence of dust is not always sufficient to detect young stars. The radial velocity map shows an edge-on, rotating disk in the center, confirmed by a central drop in the velocity dispersion. This central disk contains old stellar populations, probably slightly more metal rich (see the age and  metallicity maps). 

The next galaxy is also an Sab galaxy, NGC~4274 (Fig.~\ref{N4274}).  This galaxy seems not to be
very different from the one before. The unsharp masked image shows a central
spiral, which might be similar to the one in NGC~3623 (which is edge-on, so the
spiral structure in the dust cannot be seen). This spiral is associated to a
rotating feature, seen both in the velocity and the velocity dispersion maps.
In both galaxies, ionized gas is present everywhere in the central regions (Falc\'on-Barroso et al. 2006).
Different from NGC~3623, the stellar populations in the central disk in NGC~4274 are much
younger than in the main body of the galaxy, as can be seen from the line
strength maps, especially H$\beta$ absorption. If one now looks at the
photometric decomposition (Peletier 2009), one sees that the part of the 
surface brightness profile of NGC~4274 that lies above the large exponential
disk, i.e. the bulge, corresponds to the region of the inner disk, and is best
fitted by a S\'ersic profile with $n$ = 1.3. In the case of NGC~3623 the
so-defined bulge is much larger, and has a S\'ersic index of $n$ = 3.4. 
Kormendy \& Kennicutt (2004) would call the bulge in NGC~4274 a pseudo-bulge,
and the one in NGC~3623 a classical bulge. However, the comparison here shows
that both objects are very similar, but that only the inner disk to elliptical bulge
ratio in both galaxies is different. 
The study of other bulges in this sample shows that many early-type spiral galaxies contain central disks, with often young stellar populations in them.

The study of the central stellar populations and dust can also be done very well using IRAC on the Spitzer Space Telescope. van der Wolk (2011) in his PhD Thesis presents color maps of these SAURON-selected spirals. Information about the ages of the stellar populations comes from the [3.6] -- [4.5] maps (see Section \ref{spitzercol}). In Fig.\ref{wolk} the [3.6] - [8.0] maps of the 2 galaxies are shown. These maps show the amount of warm dust (mainly due to PAHs). One can see that in NGC~3623 relatively little warm dust is present, consistent with the absorption line maps, while much more is present in NGC~4274. 
\smallskip

\begin{figure}
\begin{center}   
\includegraphics[width=\textwidth,clip]{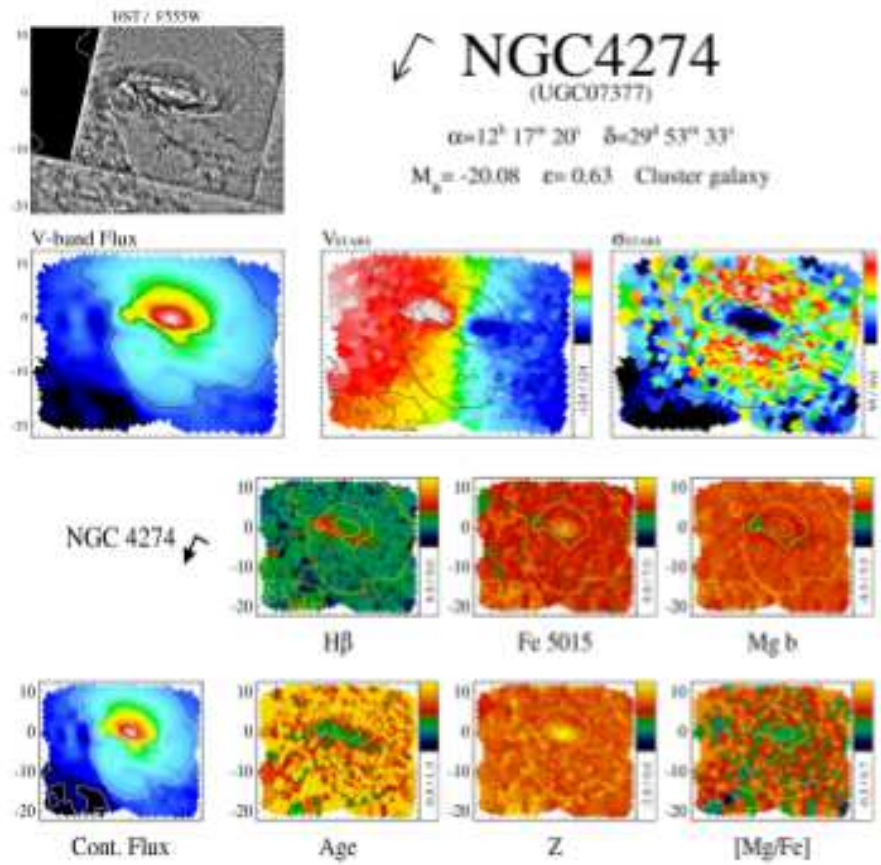}
 \caption{Same as Fig.\ref{N3623}, but now for NGC~4274.}
\label{n4274}
\end{center}
\end{figure}

\begin{figure}
\begin{center}   
\includegraphics[width=\textwidth,clip]{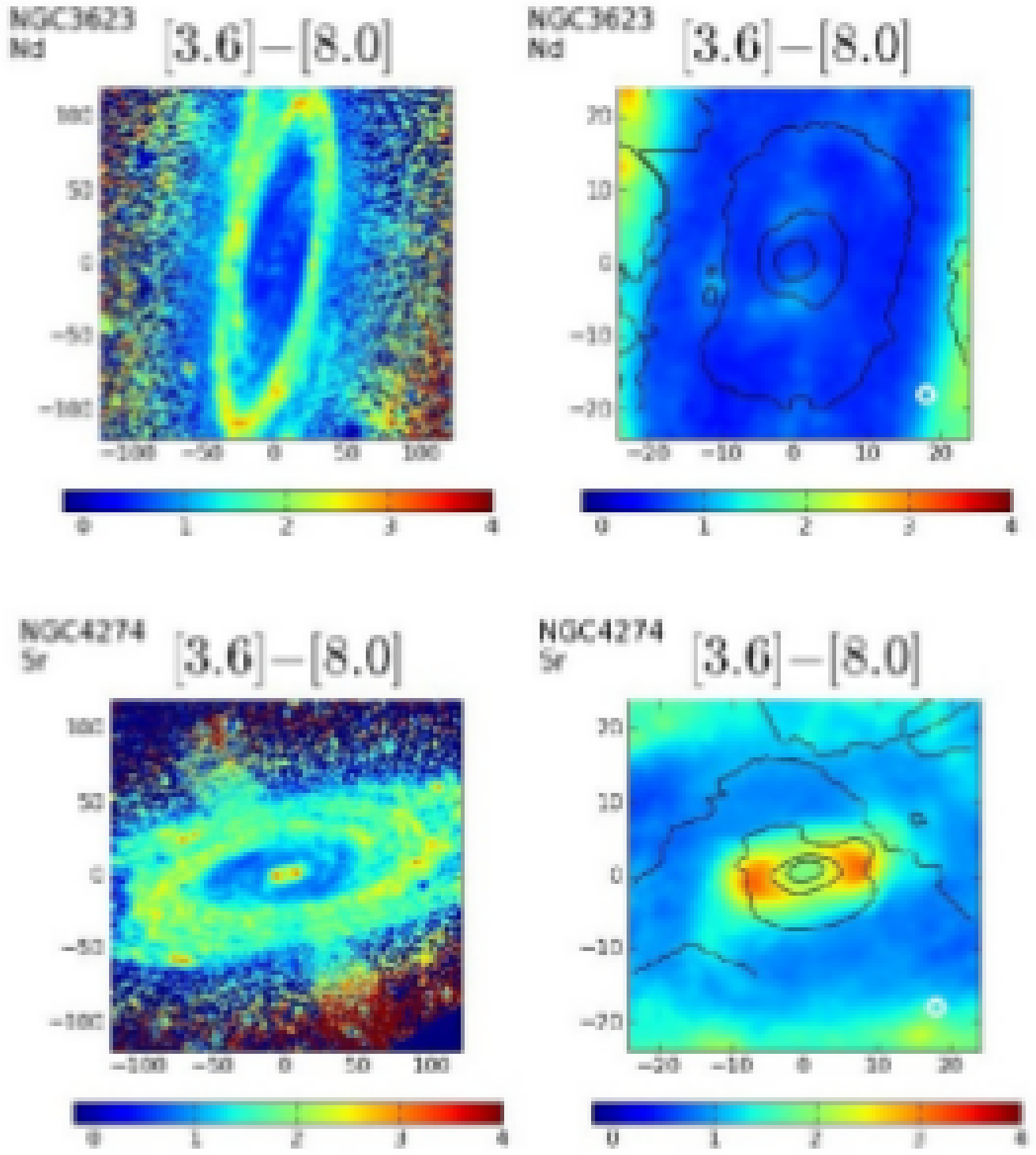}
 \caption{Spitzer IRAC [3.6] - [8.0] color maps of the spiral galaxies NGC~3623 and NGC~4273 (from van der Wolk 2011). Shown are color maps of the whole galaxy, and a central zoom. This color, a good indicator of warm dust, shows a considerable amount of dust in the center of NGC~4274, corresponding to the region of the inner spiral in Fig.~\ref{n4274} and  much less dust in the center of NGC~3623 (Fig.\ref{N3623}).}
\label{wolk}
\end{center}
\end{figure}

\subsection{Stellar masses and the IMF in galaxies}
\label{IMFMass}

Spectra and colors of SSPs are fairly insensitive to the initial mass function (IMF), because most of the
light comes from stars in a narrow mass interval around the mass of stars at the main sequence turnoff. On
one hand this is good, because it allows modelers to produce predictions for the spectra of galaxies that
are accurate at most wavelengths. However, the same effect makes it possible to hide a large amount of mass
in the form of low mass stars  in a stellar population, making the stellar mass-to-light ratio a badly
constrained parameter. Colors and lines of galaxies can generally be fitted well with a Salpeter IMF (a
power law function with $x$=1.3, see above). However, the same observables can also be fitted with an IMF
that flattens below a certain critical mass, e.g. the Chabrier (2003) IMF, which flattens off below 0.6
M$_\odot$, giving a M/L ratio which is a factor 2 lower.

Until the end of the 1990's the uncertainties in the IMF were considered so important that estimates of
stellar mass were rarely given. This changed with the influential paper by Bell \& de Jong (2001), who
showed that if one maximized the stellar mass in the disk when reproducing rotation curves of galaxies (the
so-called maximum disk hypothesis) an IMF similar to the Salpeter IMF at the high-mass end with fewer
low-mass stars, giving stellar M/L ratios  30\% lower than the Salpeter value, was preferred. After this,
it has become very common that stellar masses are given when fitting lines or colors of galaxies. In the
important paper of Kauffmann et al. (2003), where stellar masses of many galaxies in the SDSS survey are
calculated, the Kroupa (2001) IMF is used, a similar kind of IMF, and in only 2 sentences the authors
mention that there can be systematic uncertainties in the derived stellar masses, as a result of the choice of the
IMF.

\begin{figure}
\begin{center}   
\includegraphics[height=10cm,clip]{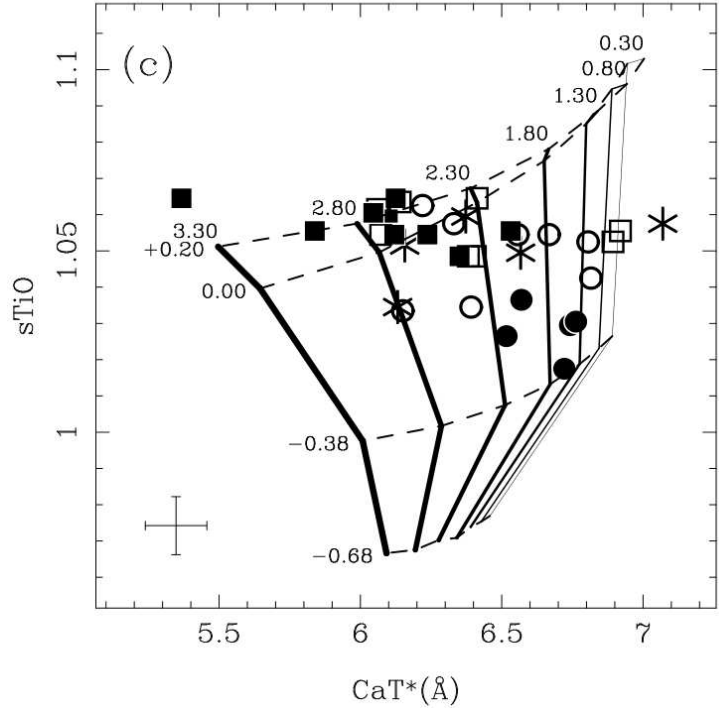}
 \label{cenarro2003}
\caption{SSPs model predictions at fixed old age with varying power-like
IMF slopes (x = 0.3 --3.3, see the labels) and metallicity from --0.68 to 0.20. Different symbols indicate galaxies with different velocity dispersions (see Cenarro et al. 2003).} 
\end{center}
\end{figure}

Although in the optical most features are only slightly sensitive to the IMF-slope, there are some, mainly
in the infrared, which strongly depend on the dwarf to giant ratio, i.e. the IMF-slope. Examples are the
Wing-Ford band at 0.99 $\mu$m, the Na I doublet at 8190 \AA, and the Ca II IR triplet around 8600\AA. These
lines have been used by several authors to constrain the IMF-slope (e.g., Spinrad \& Taylor 1971, Faber \&
French 1980, Carter et al. 1986, Schiavon et al. 2000, Cenarro et al. 2003), but the results have never
been very conclusive. The most important reason for this is that telluric absorption lines make it hard to
measure accurate line strengths in this region of the spectrum. The second reason is that it is not always
straightforward to derive the IMF-slope from the observations.  For example, in Fig.\ref{cenarro2003}
Cenarro et al. showed, based on the strength of the Ca II IR triplet and a molecular TiO index, and on the
anti-correlation between the strength of the Ca II IR triplet and the velocity dispersion, that the IMF
slope in elliptical galaxies increases for larger galaxies. However, other solutions are possible, e.g.,
that the [Ca/Fe] abundance ratio becomes lower for more massive galaxies. One might also think about
systematic errors in the stellar population models that are needed to establish the IMF-slope. For example,
in the models that Cenarro et al. use, solar abundance ratios are used in the stellar evolutionary
calculations, and the stellar library used mostly consists of stars in the solar neighborhood, which
implies that here too the abundance ratios must be close to solar.

Recently, van Dokkum \& Conroy (2010, 2011), and Conroy \& van Dokkum (2012) have revived this topic. Using new
methods to better remove the atmospheric absorption lines and new models in the near-infrared, they present
conclusions that the IMF-slope increases with increasing galaxy velocity dispersion (mass). For the
largest galaxies the IMFs found are a bit steeper than Salpeter (x=1.6). This implies that stellar masses
inferred from stellar population analysis will have to be increased by a small factor, which will not be
larger than 2.  Although this is an important result, one should remember the caveat that such a result
depends on the stellar population models, which for non-solar abundance ratios are not perfect yet. 
Similar remarks can be made about the recent paper of Ferreras et al. (2012), who confirm Conroy \& van
Dokkum's result using the Na doublet at 8200\AA\ with stacked data of a large sample of SDSS galaxies, and
of Smith et al. (2012) who use an area near the Wing-Ford band. Recently, Cappellari et al. (2012) claim
that independent analysis based on stellar dynamical fits to 2-D kinematic fits to galaxies of the Atlas-3d
survey confirms the IMF trends observed in stellar populations.

If, when calculating stellar masses, one still doesn't want to depend on these estimates of the IMF slope,
one can also use photometry or indices further to the infrared. For example, it is known that M/L ratios in
the K-band vary little with stellar populations, since here the relative contribution to the light of
dwarfs vs. giants is much smaller than in the optical. The same holds for the Spitzer [3.6] and [4.5]
bands, which are still dominated by light from stars.  Meidt et al. (2012) nicely show how stellar masses
can be obtained from images in both these bands for spiral galaxies.

\subsection{Beyond the optical}
\label{uvnir}

Stellar population synthesis in the UV is less well developed, because of various reasons. First of all, the amount of data available is limited, since it all has to come from space. Secondly, interpretation is complicated, since a few hot stars can over-shine all other stars, making it very difficult to obtain information on the not so young stellar populations.  

\begin{figure}
\begin{center}   
\includegraphics[height=10cm,clip]{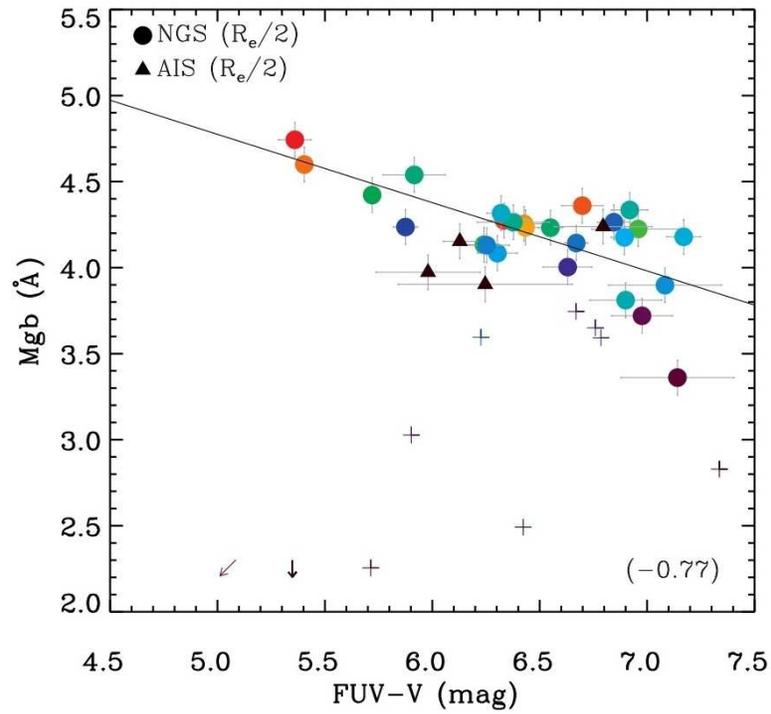}
 \caption{Updated FUV -- V vs. Mg~b diagram (also called Burstein diagram), 
 from Bureau et al. (2011). Galaxies with H$\beta$ $>$ 1.8 \AA\ are indicated with  'plus'-signs.}
\label{bureau2011} 
\end{center}
\end{figure}

Burstein et al. (1988) published a large number of IUE-spectra of early-type galaxies. Their main result was a relation between the 1550 -- V color (1550 is here a passband with effective wavelength 1550\AA\ ) and the optical 
Mg$_2$ index. Massive galaxies with large Mg$_2$ index have a very blue 1550 -- V color. This effect, the so-called UV-upturn, is probably due to extreme horizontal branch stars, but can also have other reasons (see O'Connell 1999 and Yi 2008 for reviews). With new and better quality GALEX-data, Bureau et al. (2011) show that this effect is present only when the optical H$\beta$ index is low, which implies that from the optical spectrum there is no evidence for any young stellar populations (see Fig.~\ref{bureau2011}.

Very little has been done on the analysis of line strength indices. This is surprising, since the UV is particularly important for the analysis of high redshift spectra. Recently, Maraston et al. (2009) published some stellar population models based on the IUE-stellar library of Fanelli et al. (1992).
A problem with this empirical library is, that its range in metallicity is small. However, in this region empirical stars are probably more reliable than synthetic spectra, due to difficulties treating the effect of stellar winds that affect the photospheric lines of massive stars. There are still considerable differences between using the Fanelli library and a high-resolution version of the Kurucz library of stellar spectra (Rodr\'\i guez-Merino et al. 2005) in the Maraston models.  

There are ongoing efforts to develop a stellar library from HST/STIS stars, providing higher S/N and higher resolution spectra than IUE covering a much larger parameter space (the NGSL library - Gregg et al. 2006). This library has not been incorporated into any stellar population models yet, although it has been characterized and stellar parameters have been homogenized in Koleva \& Vazdekis (2012). 

Just like the UV, the near-IR has also not been studied very much. While
broadband colors are predicted by many stellar population models,
very few spectrophotometric models are available. The problem has been mentioned before. The NIR is dominated by evolved stellar populations, i.e., RGB and especially AGB, of which the number and lifetimes are not well known, since they are so short-lived that good statistics cannot be obtained from globular and open cluster HR-diagrams. Furthermore, AGB stars lose large amounts of mass, making their lifetimes and also their spectra uncertain. On top of that, they are highly variable. 
Spectrophotometric models at a resolution of $\sim$ 1100 are available from  Mouhcine \& Lan\c con (2002). They are based on about 100 observed stars from Lan\c con \& Wood (2000) for static luminous
red stars, stars from Lan\c con \& Mouhcine (2002) for oxygen rich
and carbon rich LPVs, and the theoretical library of Lejeune et al.
(1997, based on Kurucz models) in all other cases. Conroy \& van Dokkum (2012) recently made some models using the IRTF library (Rayner et al. 2009, Cushing et al. 2005). 
At low resolution ($\sim$ 50\AA\ ), there are models from Maraston (2005) and Charlot \& Bruzual (Version of 2007, unpublished), based on theoretical atmospheres, and only tested in the broad bands J, H and K. Maraston also  presents some low resolution indices.
The problem with the models at present is that only the broad band  fluxes have been tested well using clusters and galaxies, but that detailed testing of line indices or narrow band fluxes is still lacking. For example,  
in a recent paper, Lyubenova et al. (2012) showed that globular clusters cannot be fitted by the models of Maraston (2005) models in the C$_2$ - D$_{CO}$ diagram. 
C$_2$ indicates the line strength of a feature at 1.77 $\mu$m (Maraston 2005), while D$_{CO}$ is an index measuring the strength of the CO band head at 2.29 $\mu$m (M\'armol-Queralto et al. 2008). 
The problem here is a lack of Carbon stars in the models, stars of $\sim$ 1 Gyr (Lan\c con et al. 1999). It indicates that making models of the TP-AGB phase is very difficult (see also Marigo 2008). 
The situation might be improving soon. 
Better data of nearby galaxies and clusters are becoming available (e.g. Lyubenova et al. 2012, Silva et
al. 2008, M\'armol-Queralto et al. 2009). And better stellar libraries are expected (e.g., the X-Shooter
library (Chen et al. 2011)). By comparing data with models, we will learn where the models should be improved, up to the moment that the NIR will give useful constraints to galaxy evolution theories.

\subsection{Stellar population analysis from surface brightness fluctuations}
\label{sbf}

\begin{figure}
\begin{center}   
\includegraphics[width=\textwidth,clip]{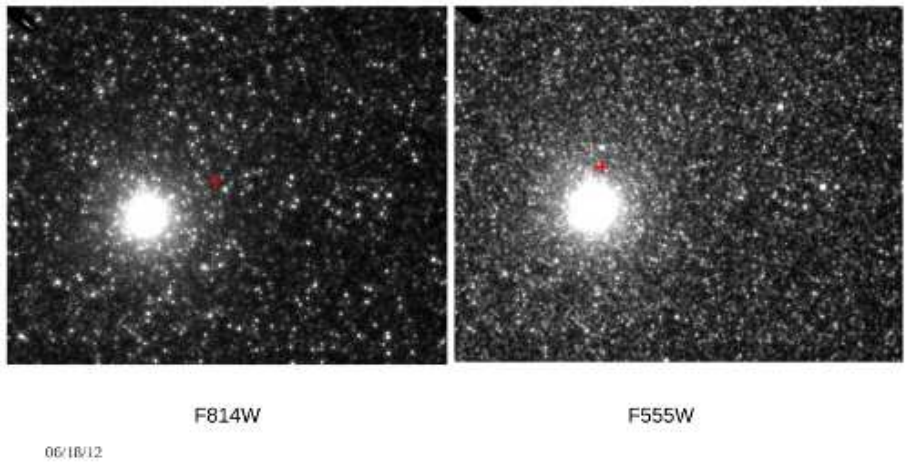}
 \caption{The effect of surface brightness fluctuations: ACS images of NGC 205 in F814W (left) and F555W (right).}
\label{sbffig}
\end{center}
\end{figure}

In Fig.\ref{sbf} one sees the nearby dwarf elliptical NGC 205 in 2 bands. In the redder band, F814W, the
giants can be distinguished much more easily from the underlying mass of fainter stars than in F555W. One
can imagine that if this galaxy is placed at larger distances, one can see the individual giants up to
larger distances in F814W. One can use the noise map, obtained after removing a smooth model of the galaxy,
as a measure of the galaxy distance. Even more, since this noise map strongly depends on the number of
bright giants and supergiants, one can use the noise characteristics, or the {\it surface brightness
fluctuations} as a way to characterize the stellar populations in a galaxy.

A review about surface brightness fluctuations as a stellar population indicator is given in Blakeslee
(2009). It shows that the method can be used well for determining distances in early-type galaxies (giants
or dwarfs), but that the use for stellar population analysis is still limited to the optical. In the
near-IR there are considerable discrepancies between surface brightness fluctuations predicted by models
and the observations (Lee et al. 2009). With the advent of new, large telescopes, this work will
undoubtedly become more important in the future.

\section{The use of scaling relations}
\label{scalrel}

Rather than studying individual galaxies one can also study them by investigating the 
evolution of so-called scaling relations. Nearby galaxies happen to display clear correlations between well-defined and
easily measurable galaxy properties. With high redshift studies
now routine, scaling relations are more useful than ever, allowing
us to probe the evolution of galaxy populations over a large
range of lookback times (e.g. Bell et al. 2004, Saglia et al. 2010).
In this review I will discuss the color -- (and line strength --) $\sigma$ relation, a potentially tight
relation connecting the galaxy mass to its stellar populations, and the fundamental plane of galaxies, a
relation connecting the structure of galaxies to their mass.

\subsection{The color -- $\sigma$ relation}
\label{colsig}

It has been known now for more than 50 years that early-type galaxies show a tight color-magnitude (C-M) relation (Baum 1959, Sandage
1972; Visvanathan \& Sandage 1977), in the sense that larger galaxies are redder. Bower et al. (1992) showed that in the Virgo and Coma clusters, when taking small central apertures, the scatter in $U-V$, and $V-K$ is extremely small. Noting that when galaxies age, their color becomes redder, and their velocity dispersion only changes slightly, they could show that  cluster ellipticals are made of very old
stars, with the bulk of them having formed at z $>$  2. so, by assuming that for a given $\sigma$ the eldest galaxies are situated at the reddest color, and that along the color -- $\sigma$ relation the metallicity (and possibly also the maximum age for galaxies of a certain $\sigma$) changes, one can derive relative ages and metallicities using color -- $\sigma$ relations. It has been used as an important
benchmark for theories of galaxy formation and evolution 
(e.g. Bell et al. 2004; Bernardi et al. 2005)
Galaxies devoid of star formation are thought
to populate the red sequence, while star-forming galaxies lie in the
blue cloud (e.g. Baldry et al. 2004). The dichotomy in the distribution
of galaxies in this relation has opened a very productive	
avenue of research to unravel the epoch of galaxy assembly (e.g.
De Lucia et al. 2004; Andreon 2006; Arnouts et al. 2007).
 
This  stellar population --  mass relation for galaxies has manifested itself in the literature in
many flavors. Various colors have been used, from blue colors that are very age-sensitive to red
colors covering a large wavelength baseline (e.g. V-K). To avoid the effects of dust extinction, one
often uses line strengths instead of colors. The Mg$_2$ -- $\sigma$ relation has been used very
frequently in the literature (e.g. Terlevich et al. 1981). SDSS related studies have been
concentrating on the H$\delta$ line and the D$_{4000}$ break (Kauffmann et al. 2003), finding that
galaxies less massive that 3~10$^{10}$ M$_\odot$ are predominantly younger than more massive galaxies.
Cenarro et al. (2003) find an inverse relation of the Ca II IR triplet strength as a function of
$\sigma$, which up to recently is not well understood, and might have something to do with IMF-changes
in galaxies (see Section \ref{IMFMass}). The galaxy mass indicator (i.e., here $\sigma$) can be replaced by
other indicators such as galaxy luminosity, stellar mass, etc. When using stellar mass, the
relations are not so tight (e.g. Peletier et al. 2012), since compact ellipticals fall off the
relation for the other galaxies. Compact ellipticals have higher $\sigma$ and also redder colors/line
indices than one would expect from their stellar mass. What helps in any case is taking into account
both random ($\sigma$) and regular motion (rotation) (Zaritsky et al. 2006). This way both ellipticals
and spiral galaxies can be compared easily with each other (Falc\'on-Barroso et al. 2011).  An advantage is
that the color -- $\sigma$ relation is independent of galaxy distance. When using mass, or
luminosity, the errors involved in measuring these distances will generally dominate the scatter,
unless these errors can be avoided, in e.g. a galaxy cluster.

\begin{figure}
\begin{center}   
\includegraphics[width=\textwidth,clip]{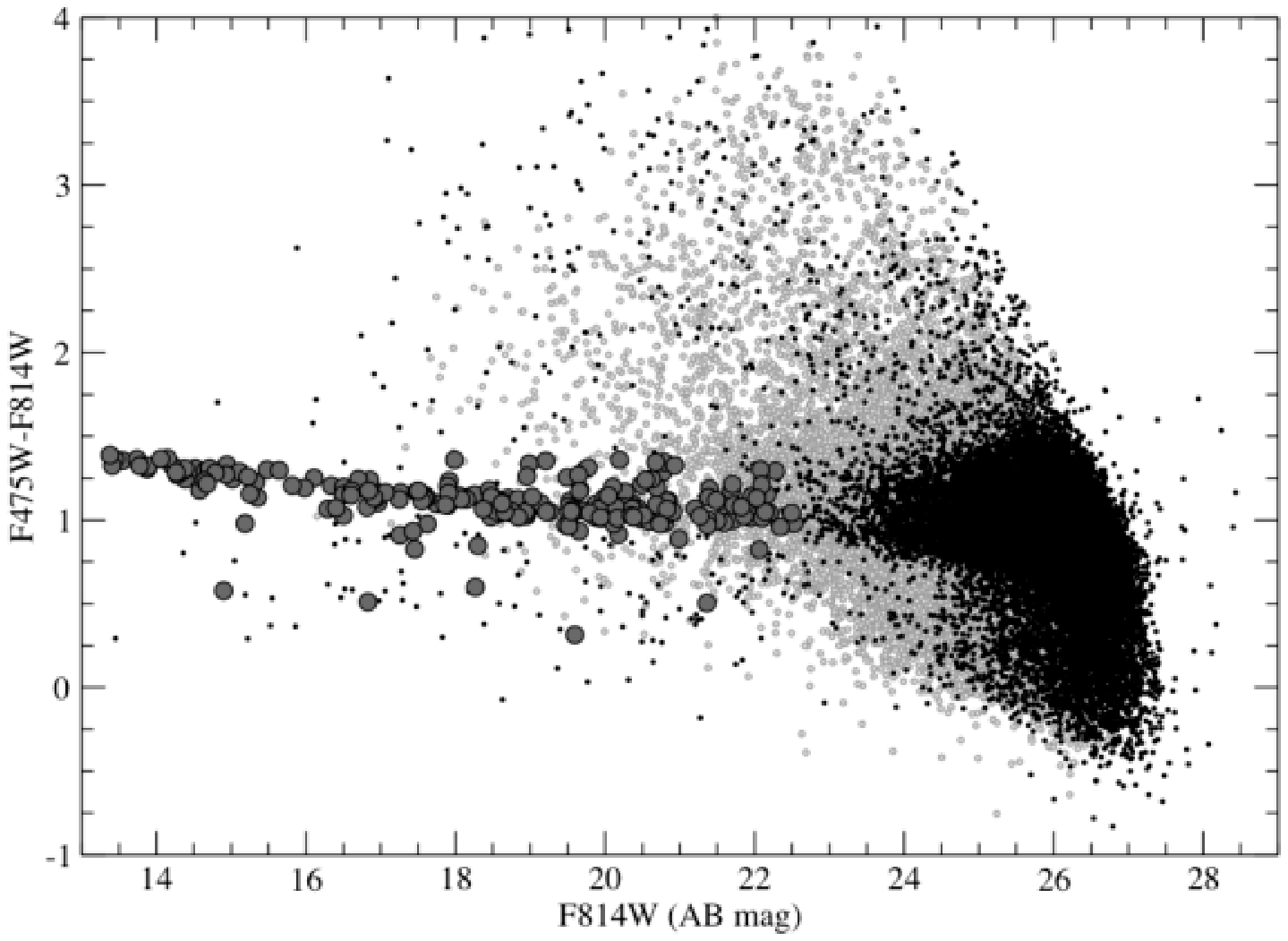}
\includegraphics[width=\textwidth,clip]{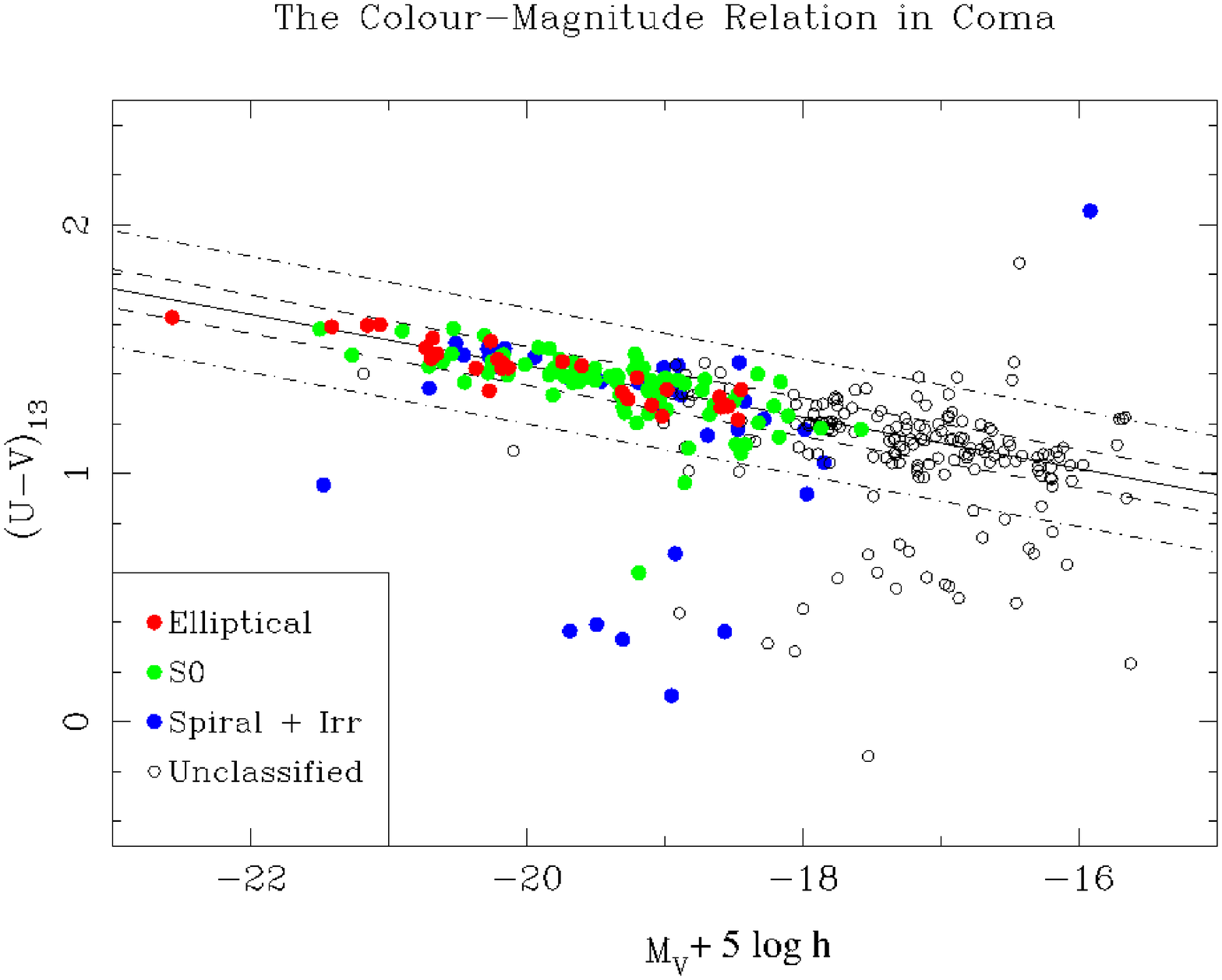}
 \caption{The color - magnitude relation in the Coma cluster. Top: $F475W -  F814W$ vs. F814W relation from the Coma-ACS survey (Hammmer et al. 2010). These apparent magnitudes can be converted to absolute magnitudes using $m-M ~=~ 35$. Bottom: $U-V$ vs. M$_V$ relation from Bower et al. (1999).  }
\label{colsigfig}
\end{center}
\end{figure}

In Fig.~\ref{colsigfig} I show two color -- magnitude relations in the Coma cluster. On the right is
shown the result of Bower et al. (1999), showing spectroscopically confirmed cluster members.
One sees that ellipticals and S0s form a tight color -- magnitude relation. Spiral galaxies are bluer
for a given magnitude, indicating younger ages. At fainter magnitudes more and more galaxies are falling blue-ward of the
relation, showing that star formation in smaller galaxies is more common. On the left, a diagram is
shown from the Coma-ACS survey (Hammer et al. 2010) a survey at a much higher
resolution of a small part of the Coma cluster. Many dwarfs are included. At faint magnitudes, many
galaxy are shown to be redder or bluer than the linear relation. These are mostly background galaxies,
shown by the fact that the larger symbols, the spectroscopically confirmed Coma cluster members,
almost all lie on or below the relation. A few compact ellipticals in Coma lie above the relation.

\begin{figure}
\begin{center}   
\includegraphics[width=\textwidth,clip]{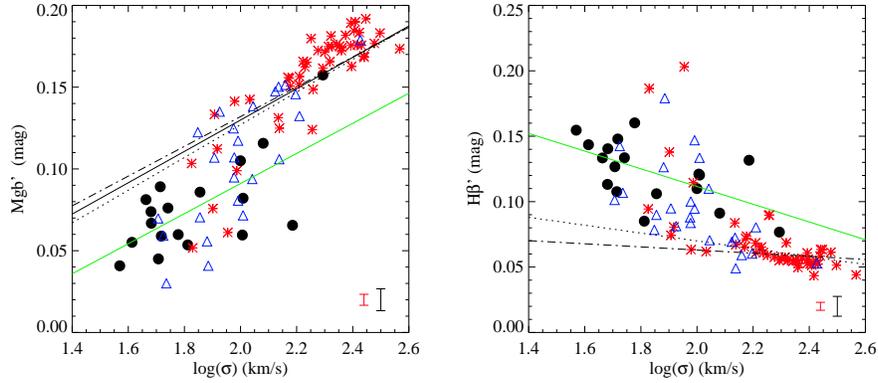}
 \caption{Mg$~b$ -- $\sigma$ and H$\beta$ -- $\sigma$ relations for 90 ellipticals and bulges from the SAURON sample (from Ganda et al. 2007). Elliptical galaxies are indicated with red asterisks, Sa's with blue open triangles, and later type spirals with filled black circles.}
\label{ganda}
\end{center}
\end{figure}

Also for spiral galaxies, the color -- $\sigma$ relation can be used very well to study stellar
populations. Falc\'on-Barroso et al. (2002) showed that bulges with old stellar populations fall on the
tight Mg~b -- $\sigma$ relation for elliptical galaxies and S0's. This means that the stellar populations
in the bulge of a galaxy is not determined by the mass of the whole galaxy, but by the mass (or $\sigma$)
of the bulge (similar to the black hole mass - Ferrarese \& Merritt 2000). So, by plotting an index, such
as Mg~b against central $\sigma$, one can use the relation in the same way as for ellipticals. In
Fig.~\ref{ganda} this is done for the spirals of the SAURON survey (Peletier et al. 2007, Ganda et al.
2007). Plotted are central line strengths. The figure shows that the central stellar populations in
late-type spirals are all younger than the ones in early-type galaxies. Sa's show a large scatter, and have
luminosity-weighted stellar populations that range from very young to as old as ellipticals. The diagram on
the right, which uses H$\beta$ (in magnitudes) is maybe a better diagram to use, since here the dependence
of the index on $\sigma$ is much lower, which means that one can read off ages much more easily. The
H$\beta$ -- $\sigma$ diagram has not been used very often in the literature, since only recently one is
able to clean the absorption from the H$\beta$ emission. The H$\beta$ -- $\sigma$ diagram shows the same
results as Mg~b -- $\sigma$: much more scatter in the stellar populations in Sa's (and some small E's and
S0's) than in later type galaxies, which, on the other hand, are younger than ellipticals.

\subsection{Stellar Population Analysis from Spitzer Colors}
\label{spitzercol}

\begin{figure}
\begin{center}   
\includegraphics[width=\textwidth,clip]{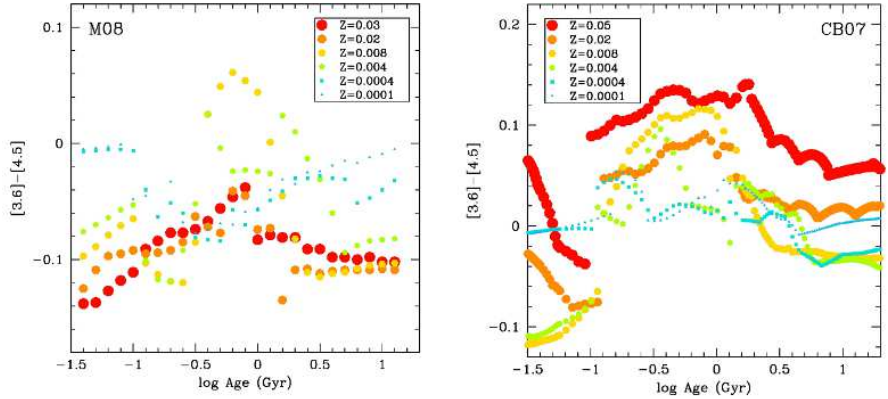}
\caption{SSP Models for [3.6] -- [4.5] by Marigo etal. (2008, left) and 
Charlot \& Bruzual (Version of 2007, unpublished) as a function
of age and metallicity. }
\label{spitzermodels}
\end{center}
\end{figure}

In recent years the Spitzer Space Telescope has made many observations in 4 bands. The shortest wavelength
bands, [3.6] and [4.5], in nearby galaxies  are mainly dominated by stellar light, while the [8.0] bands
mainly detects warm dust from particles like PAHs (Fazio et al. 2004). Since the light at 3.6 and 4.5
$\mu$m is barely affected by extinction, and also not by young, hot stars, the [3.6] - [4.5] color seems
to be useful to study the cold stars in early-type galaxies. The color will be affected by AGN and TP-AGB
stars, and will also be dependent on metallicity. In Fig.\ref{spitzermodels}  predictions are shown for the
[3.6] -- [4.5] color from Marigo et al. (2008) and Charlot \& Bruzual (Version of 2007, unpublished) for SSPs of various ages and
metallicity. Note that in both sets of models stellar populations of $\sim$ 1 Gyr make this color
particularly red ([3.6] -- [4.5] goes to larger values). The models do not agree with each other in
predicting the dependence of [3.6] -- [4.5] as a function of metallicity for old ages: in the Marigo (2008)
models [3.6] -- [4.5] becomes bluer with increasing metallicity, while in Charlot \& Bruzual the colors
becomes redder. 

\begin{figure}
\begin{center}   
\includegraphics[height=\textheight,clip]{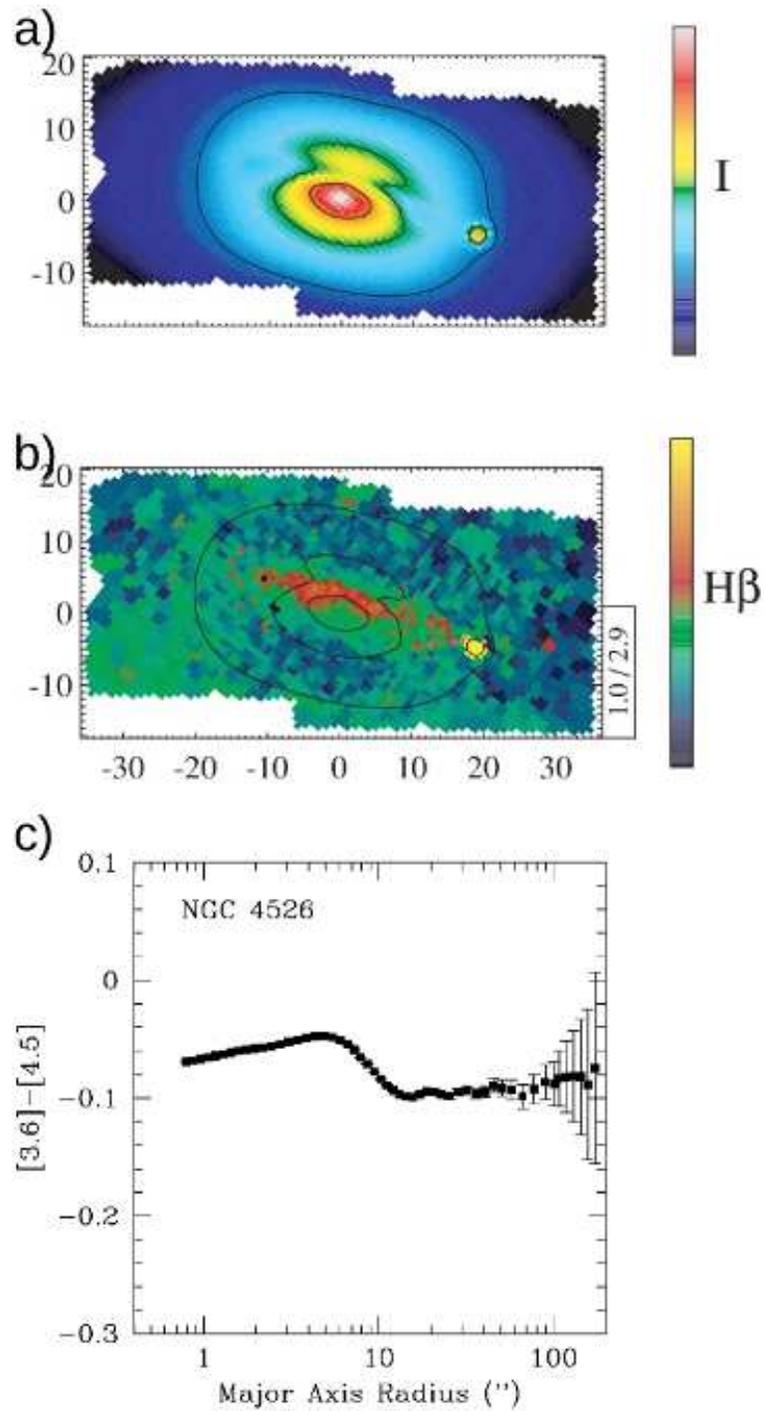}
 \caption{SAURON images in V-band continuum (a)) and H$\beta$
absorption for NGC~4526. The yellow spot is a foreground
star. In the bottom panel is shown the [3.6] - [4.5] profile. The central
redder part of the profile corresponds exactly to the inner disk
seen in the upper 2 panels (from Peletier et al. 2012).}
\label{n4526composite}
\end{center}
\end{figure}

In Peletier et al. (2012) we describe a study of the [3.6] -- [4.5] color in the 48 early-type galaxies of
the SAURON sample. It is shown that the images in the 2 bands look like smooth, elliptical galaxies in the
optical, without dust lanes etc. For every object colors were determined in circular apertures of r$_e$
and r$_e$/8. Also radial color profiles were determined by {\bf 1.} Convolving the 3.6 $\mu$m image with
the 4.5 $\mu$m PSF and vice-versa, to remove any PSF-effects near the center; {\bf 2.} Fitting the same
ellipses, with fixed center, ellipticity and position angle in both bands; {\bf 3.} Performing accurate sky
subtraction; and {\bf 4.} Making the ratio of both profiles. One color profile is shown in
Fig.\ref{n4526composite}, where we show the (optical) SAURON continuum image, the H$\beta$ absorption map,
and the [3.6] -- [4.5] color profile. The high values in the H$\beta$ map show the stellar populations in
the dust lane, which are younger than in the main galaxy. The [3.6] -- [4.5] profile shows that these young
stellar populations make the color redder. On top of that, the general gradient is making the galaxy
slowly redder when going outwards. Given the fact that most galaxies become less metal rich going outward,
this might mean that [3.6] -- [4.5] becomes redder for decreasing metallicity, or bluer for increasing
metallicity. We can understand this when we know that the 4.5 $\mu$m band contains a large CO absorption
band. When the metallicity increases, this band gets stronger, making [3.6] -- [4.5] bluer (see Peletier et
al. 2012).

\begin{figure}
\begin{center}   
\includegraphics[width=\textwidth,clip]{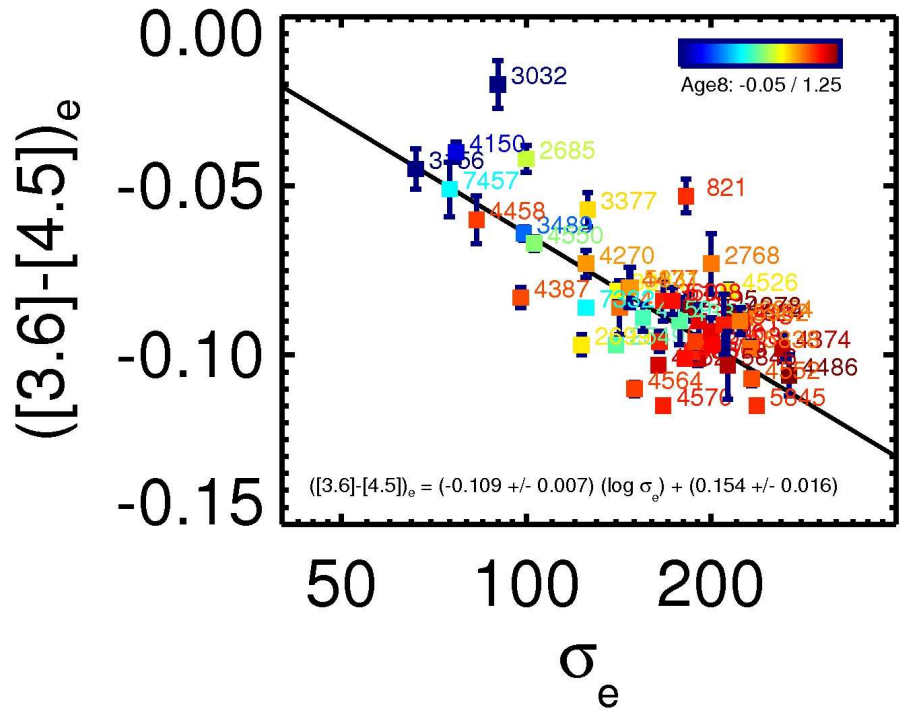}
 \caption{[3.6] -- [4.5] color as a function of velocity dispersion in km/s. The velocity dispersion has been measured within r$_e$. Here
the color, determined within 1 effective radius, is shown.}
\label{colsig_age}
\end{center}
\end{figure}

Most galaxies have colors everywhere between -0.15 and 0. An exception is M~87, the central Virgo galaxy, which has a very red center, due to the synchrotron emission in center and jet. No other galaxies contain such prominent central point sources. When plotting the relation of [3.6] -- [4.5] and $\sigma$ (Fig.\ref{colsig_age}) we see that both quantities are strongly related. In this diagram we have colored the galaxies with their age inside r$_e$/8, as obtained by Kuntschner et al. (2010) from the SAURON line indices. Had we used ages within r$_e$, the figure would have
been similar, but with a smaller range in colors. This is
because in these early-type galaxies many more young features
are seen in the inner parts than further out. 
The color -- $\sigma$ relation shows that more massive galaxies
are bluer. The color coding of the figure shows that these galaxies are at
the same time older, if one considers the luminosity-weighted
SSP-ages. The main difference with other colors is that the [3.6] -- [4.5]
color becomes {\sl bluer} for increasing galaxy mass/luminosity. 

So, what is the origin of this color -- $\sigma$ relation? Here one has to use mainly empirical arguments, since the models still are rather uncertain. One could think that metallicity is the main driver, with galaxies becoming less metal rich for decreasing $\sigma$, and as a result redder. On the other hand, one does not know what the metallicity dependence of [3.6] -- [4.5] is. One could also think that age is the dominant driver. In this case the fraction of AGB stars has to increase with decreasing $\sigma$. Since these stars are red, the galaxy colors then become redder. If this proves to be true, this would be a promising way to determine the contribution from AGB stars in galaxies. 
If the scatter in the color -- $\sigma$ relation can be explained by
young stellar populations on top of a much older underlying
stellar population, one would expect the outliers of the
optical line strength -- $\sigma$  relations of Kuntschner et al. (2006) to be the same as the outliers of the color -- $\sigma$  relation here. A close look teaches us that this is to first order the case. Also, there is a strong correlation between Mg~b and [3.6] -- [4.5]. If the color -- $\sigma$ relation is driven by age, it would mean that the young populations that are responsible for the bluing of [3.6] -- [4.5] are also responsible both for the decreasing Mg~b and increasing H$\beta$ index. Although it is hard to quantify what kind of SSP would be needed, the strong correlation between Mg~b, which is sensitive for stellar populations from 10$^{6-7}$ y, and [3.6] -- [4.5], which is mostly sensitive to stars above 3~10$^8$ y, would indicate that stellar populations in these galaxies would have ages older than 3~10$^8$ y. This is not very realistic, since the galaxies that are blue in the [3.6] -- [4.5] -- $\sigma$ relation always show H$\beta$ {\it emission} lines in the region of the young stars, indicating recent star formation. The alternative would be that AGB-populations are much less important than people think. That would agree with recent results from Zibetti et al. (2012), who, from near-infrared spectroscopy of post-starburst galaxies, find a much lower contribution from AGB stars than is expected from the {\it TP-AGB heavy} models of Maraston (2005). More research clearly is needed to understand the contribution of this evolved stellar population in galaxies.

\subsection{The fundamental plane of galaxies}
\label{FP}

Since its discovery (Djorgovski \& Davis 1987; Dressler et al. 1987), the
Fundamental Plane (FP) has been one of the most studied relations in the
literature. Given its tightness, like many other scaling relations the FP was
quickly envisaged as a distance estimator as well as a correlation to understand
how galaxies form and evolve (e.g. Bender, Burstein \& Faber 1992 (BBF), J\o rgensen
et al. 1996; Pahre et al. 1998; Bernardi et al. 2003). It is widely recognized
that the FP is a manifestation of the virial theorem for self-gravitating
systems averaged over space and time with physical quantities total mass,
velocity dispersion, and gravitational radius replaced by the observables mean
effective surface brightness $\mu_e$, effective (half-light) radius (r$_e$), and
stellar velocity dispersion  $\sigma$. Since velocity dispersion and surface
brightness are distance-independent quantities, contrary to effective radius, it
is common to express the FP as log(r$_e$) ~=~ $\alpha$ log($\sigma$) ~+~ $\beta$
$\mu_e$ ~+~ $\gamma$, to separate distance-errors from others. If galaxies were
homologous with constant total mass-to-light ratios, the FP would be equivalent
to the virial plane and be infinitely thin, with slopes $\alpha$ ~=~ 2 and
$\beta$ ~=~ 0.4. By studying the intrinsic scatter around the FP, one can study
how galaxy properties differ within the observed sample.

Just like the color -- $\sigma$ relation, the FP is a very useful tool to study
the evolution of stellar populations. To first order approximation, radius and
$\sigma$ are independent of stellar populations, while $\mu_e$ is. If a stellar
population ages, its luminosity decreases, and therefore also its surface
brightness. However, if one studies the evolution with redshift, one also has to
take into account the fact that galaxies become more compact with redshift
(radius evolution), and consequently their velocity dispersion increases as
well. 

\begin{figure}
\begin{center}   
\includegraphics[width=\textwidth,clip]{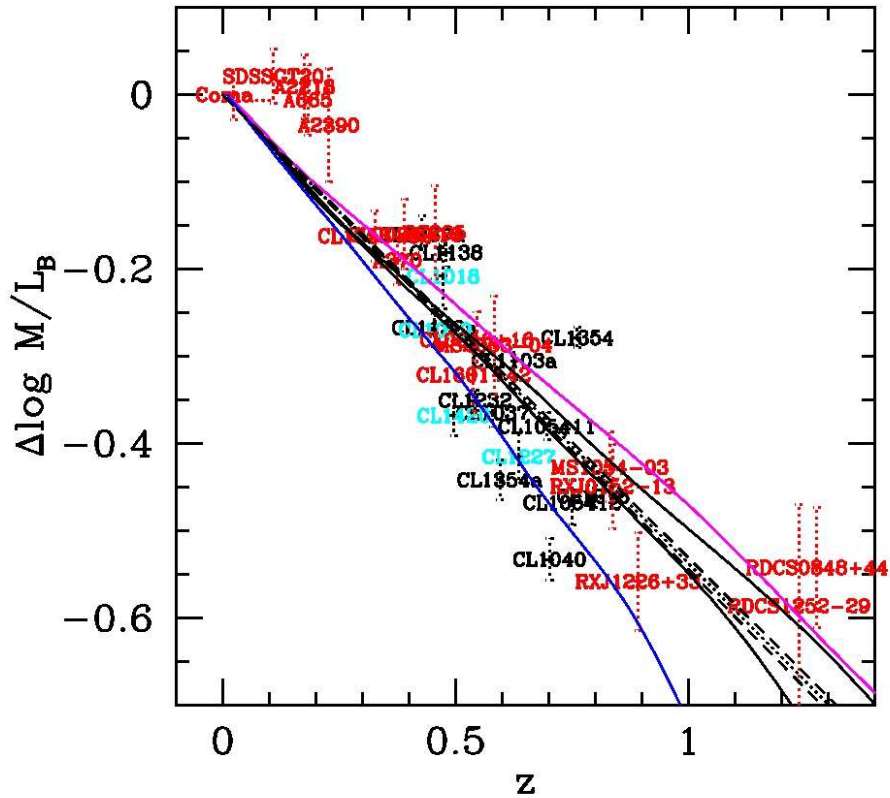}
 \caption{Redshift evolution of the $B$ band mass-to-light ratio (from Saglia et al. 2010). The full black lines show the simple
stellar population (SSP) predictions for a Salpeter IMF and formation redshift of either z$_f$ = 2
(lower) or 2.5 (upper curve) and solar metallicity from Maraston (2005). The blue line shows the SSP
for z$_f$ = 1.5 and twice-solar metallicity, the magenta line the SSP for z$_f$ = 2.5 and half-solar
metallicity. The dotted line shows the best-fit linear relation and the $\sigma$ errors dashed. }
\label{saglia}
\end{center}
\end{figure}

An important study to mention here is the EDisCS study of the FP of galaxies in
clusters up to z=0.9 (Saglia et al. 2010).  Combining structural parameters from
HST and VLT images and velocity dispersions from VLT spectra, they have been
able to determine Fundamental Plane fits for clusters with a range in redshift,
as well as for galaxies in the field. At face-value, on average, the evolution
of the surface brightness follows the predictions of simple stellar population
models with  high formation redshift ($\sim$ z~=~2) for all clusters,
independent of their total mass (see Fig.~\ref{saglia}). However, it looks as if
both the evolution of early-type galaxies with redshift and the  dependence of
this evolution on environment differ for galaxies of different mass.  These
differences manifest themselves as an evolution in the FP coefficient $\alpha$
as a function of redshift. They also find size and velocity dispersion evolution
of the sample. However, after taking into account the progenitor bias affecting
the sample (large galaxies that joined the local early-type class only recently
will progressively disappear in higher redshift samples), the effective size and
velocity dispersion evolution reduce substantially. So after making corrections
for radius and velocity dispersion evolution, they found, using SSP models,
that  massive (M $>$ 10$^{11}$   M$_\odot$) cluster galaxies are old, with
formation redshifts z $>$ 1.5. In contrast, lower mass galaxies are just 2 to 3
Gyr old.  This agrees with the EDisCS results from colors and line strength
(e.g. Poggianti et al. 2006)  who argue that the lower luminosity, lower mass
population of early-type galaxies comes in place only at later stages in
clusters. Field galaxies follow the same trend, but are  $\sim$ 1 Gyr younger at
a given redshift  and mass. This picture in principle is in agreement with the
picture one gets from stellar population analysis of nearby galaxies (Thomas et
al. 2005).

\begin{figure}
\begin{center}   
\includegraphics[width=\textwidth,clip]{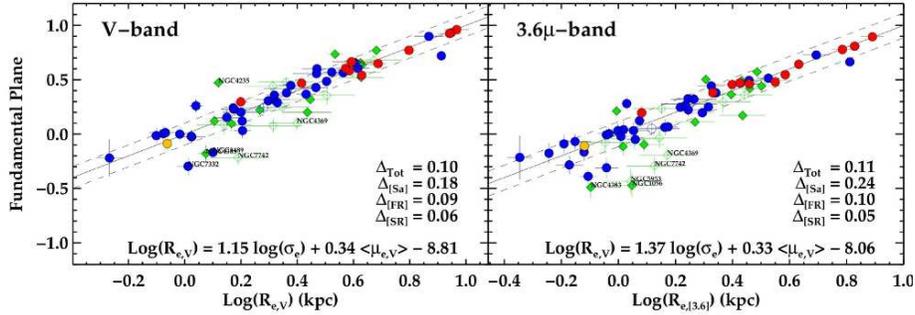}
 \caption{Edge-on views of the Fundamental Plane relation for the galaxies in the SAURON sample of galaxies in V - and 3.6$\mu$m-bands. Circles denote E/S0 galaxies, diamonds
Sa galaxies. Filled symbols indicate galaxies with good distance estimates,
open symbols those with only recession velocity determinations. In blue we
highlight Fast Rotators, in red Slow Rotators and in green the Sa galaxies.
The special case of NGC4550, with two similarly-massive counter-rotating
disc-like components, is marked in yellow. 
The solid line is the best fit relation (as indicated in the equation in
each panel) (from Falc\'on-Barroso et al. 2011).}
\label{SAURONFP}
\end{center}
\end{figure}

In the local Universe, the high S/N of the observations make it possible to look
at the FP in more detail. Here we can study the position of bulges on the FP
(e.g. Bender et al. 1992, Falc\'on-Barroso et al. 2002), the scatter in the
stellar population ages of galaxies, the amount of dark matter in various types
of galaxies along the FP, etc.  Falc\'on-Barroso et al. (2011)
(Fig.~\ref{SAURONFP}) studied the FP for the SAURON sample of 48 E/S0 galaxies and
24 Sa's. To avoid the effects of internal extinction in galaxies, they use the
Spitzer 3.6 $\mu$m band. The velocity dispersion they use is the dispersion
calculated using the integrated spectrum inside 1 effective radius. If measured
in this way, it includes both rotation and random motions (Zaritsky et al.
2006), and both ellipticals and spirals can be put on the same diagram.
Falc\'on-Barroso et al. find that the SAURON slow rotators (SR, Emsellem et al.
2007) define a very tight FP, tighter than the fast rotators. This confirms the
study from colors and line indices that SR are uniformly old systems, although
it also shows that slow rotators have the same homology (radius - surface
brightness - mass relations).  In the $V$-band the spiral galaxies deviate
because of younger stellar populations, but also because of extinction, two
effects which work in opposite directions.

\begin{figure}
\begin{center}   
\includegraphics[width=\textwidth,clip]{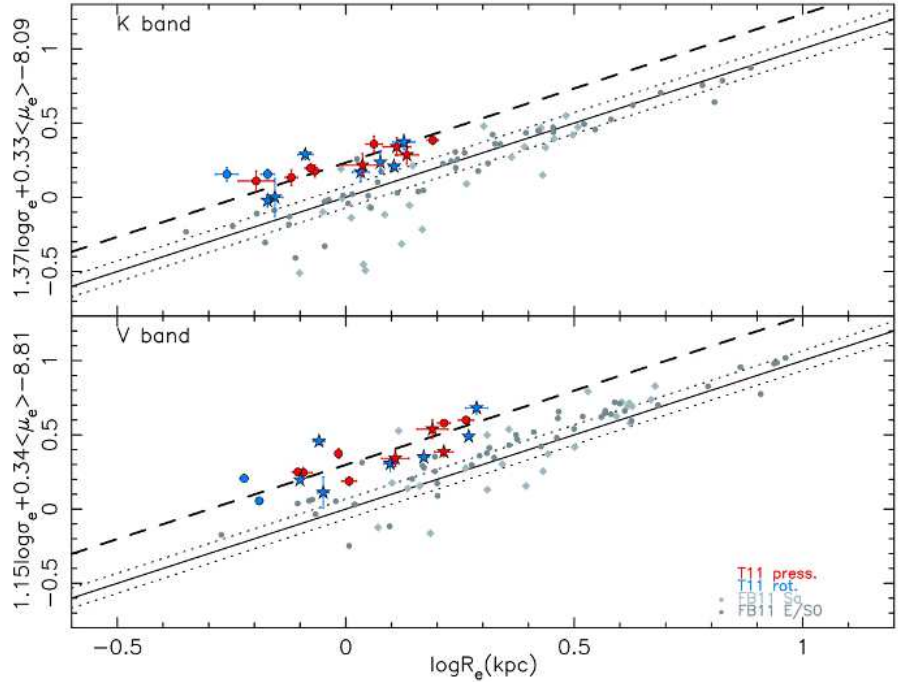}
 \caption{The position of dwarf ellipticals on FP (from Toloba et al. 2012). The early-type galaxies and Sa's are from Falc\'on-Barroso et al. 2011. Shown in red are rotationally supported dwarfs, and in blue pressure supported ones. Note that the dwarfs lie predominantly above the large galaxies, showing intrinsically higher dark matter fractions. }
\label{tolobafp}
\end{center}
\end{figure}

If one goes down in mass towards dwarf galaxies,  one traditionally finds that
dwarf ellipticals lie above the fundamental plane (BBF, de Rijcke et al.
2005). Converting the FP-parameters into new parameters $\kappa_1$, $\kappa_2$
and $\kappa_3$ using a coordinate transformation (from BBF), one can directly see how the
mass ($\propto \kappa_1$) and M/L ($\propto \kappa_3$) evolve. If one does this,
one finds that dwarf ellipticals have higher M/L ratios than ellipticals
and S0's on the fundamental plane. This result has been revised recently using
Toloba et al. (2012), who obtained high quality data for a larger sample of
dwarfs (some supported by rotation and some by random motions). From their long-slit
data they simulated the integrated spectrum inside an effective radius, to
determine the generalized dispersion (Zaritsky et al. 2006) also for the dwarfs. They
show that also the new data for dwarf galaxies fall above the fundamental plane.
Correcting for the effects of stellar populations using line indices from
Michielsen et al. (2008) in the way described by Graves \& Faber (2010)  they
find that the objects  remain above the FP, and have dynamical to stellar mass
ratios around 1.5 (see Fig.\ref{tolobafp}). If one, however, goes down to even
lower mass dwarfs, these ratios rise to much higher values (Wolf et al. 2010).

\section{Stellar Population gradients in early-type galaxies}
\label{gradients}

Metallicity gradients provide a means of studying galaxy formation. Classical
monolithic collapse scenarios (Larson 1974; Carlberg 1984) predict strong
metallicity gradients. In this scenario primordial clouds of gas sink to the
center of an over density where a rapid burst of star formation occurs. In-falling
gas mixes with enriched material freed from stars by stellar evolutionary
processes and forms a more metal rich population. Because the gas clearing time
(and hence the number of generations which enrich the interstellar medium) is
dependent on the depth of the potential well, the metallicity gradient is
dependent on the mass of the galaxy. Mergers, dominant in hierarchical galaxy
formation scenarios, will dilute existing population gradients (e.g. White 1980;
di Matteo et al. 2009), although residual central star formation can steepen
gradients again (e.g. Hopkins et al. 2009, among others) . Therefore, the study
of metallicity gradients can be used to distinguish between competing scenarios
of galaxy formation and can eventually lead to a more detailed understanding of
those scenarios. Many attempts have been made to model metallicity gradients in
more detail (e.g., Kawata \& Gibson 2003, Kobayashi 2004; Pipino et al. 2010).
They  are generally able to reproduce the gradients, but the predictive power of
the models at present is low.

\begin{figure}
\begin{center}   
\includegraphics[width=\textwidth,clip]{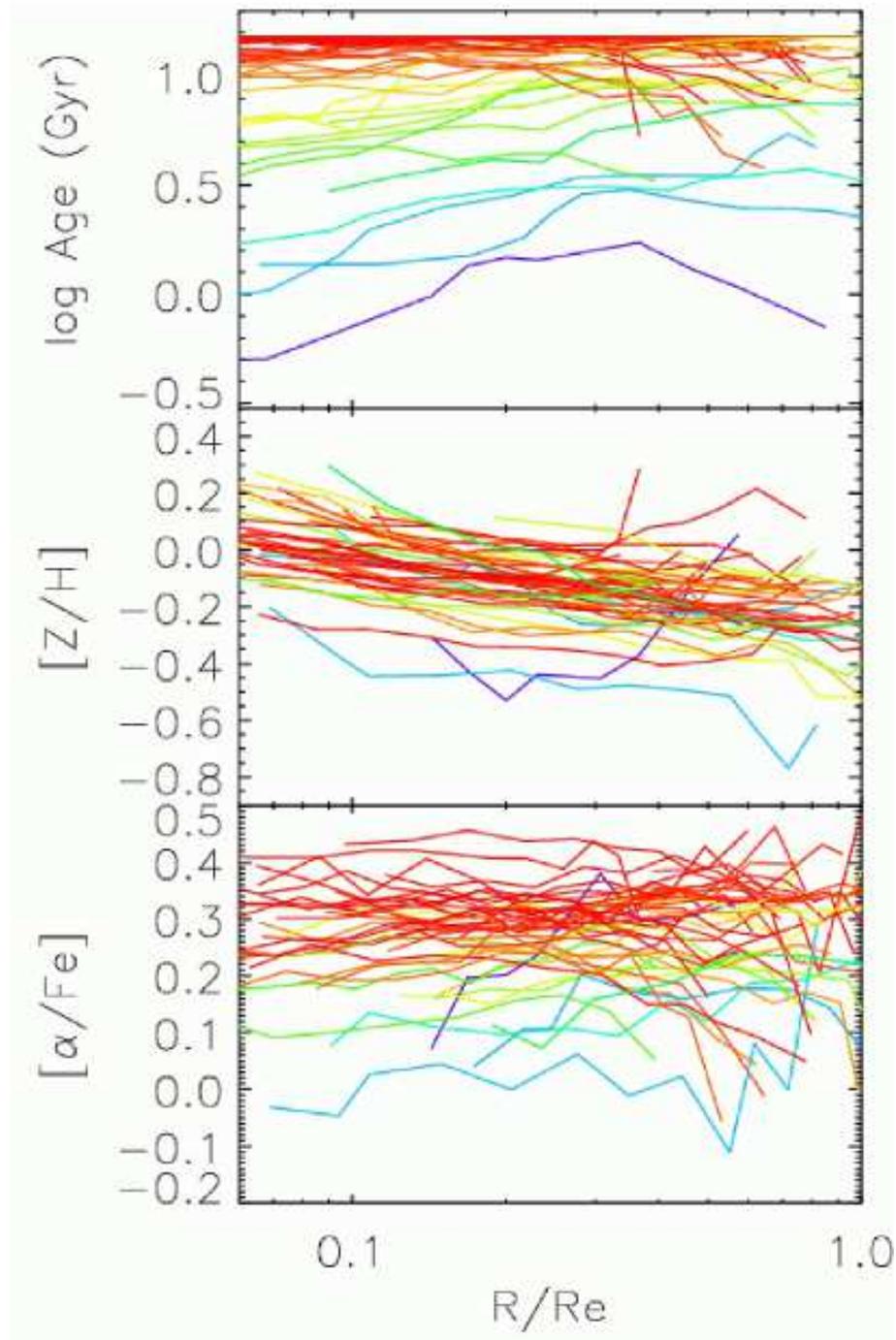}
\caption{Overview of radial profiles, averaged along isophotes, of luminosity weighted 
age, metallicity and [$\alpha$/Fe] for the 48 early-type galaxies in the
SAURON sample. The color coding reflects central r$_e$/8 SSP-equivalent
age with red referring to old stellar populations and blue to young ones (see
top plot) (from Kuntschner et al. 2010).}
\label{Kgrad}
\end{center}
\end{figure}

\begin{figure}
\begin{center}   
\includegraphics[width=\textwidth,clip]{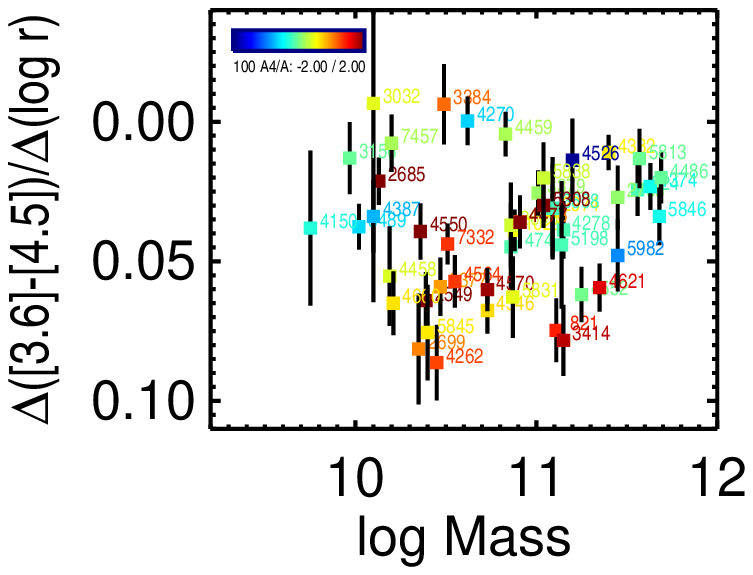}
\includegraphics[width=\textwidth,clip]{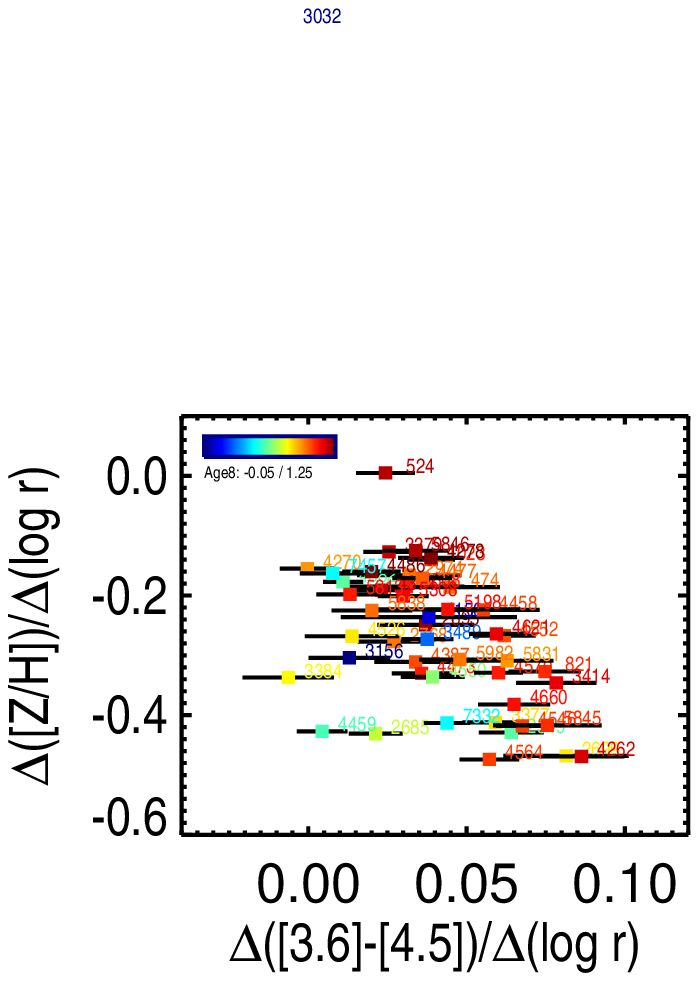}
 \caption{Top: SAURON metallicity gradients (from Kuntschner et al. 2010) as a function of the [3.6] -- [4.5] gradient (from Peletier et al. 2012). The coloring of the points is done according to the central age. Bottom: [3.6] -- [4.5] gradient as a function of mass. }
\label{Grads}
\end{center}
\end{figure}

Here I would like to discuss some new work on the gradient vs. mass relation,
which might lead to a better understanding of this relation, and therefore
galaxy formation. Using the 3 (sometimes 4) SAURON line strength indices
Kuntschner et al. (2010) have derived metallicity- and age gradients in their 48
early-type galaxies. Note that they did this assuming that at every position the
stellar populations were represented by an SSP. The resulting profiles are shown
in Fig.\ref{Kgrad}. The color of the galaxies corresponds to the
central age. One sees that galaxies become more metal poor going outward slowly,
with little difference between the gradients of the individual galaxies. This
picture is consistent with the literature (e.g. Davies et al. 1993, Carollo \&
Danziger 1993). At the same time [$\alpha$/Fe] does not change very much as a
function of radius in the galaxy. The most striking of this figure are the age
profiles. There are many galaxies which are old everywhere. These generally are
the Slow Rotators (see Section \ref{FP}), which is probably the reason their
scatter in the FP is so small. Many other galaxies are younger in the central
regions. These regions correspond to central disks, which are very common not
only in early-type galaxies, but also in spirals. Note that although many
central disks are younger, there are also central disks which are just as old as
the rest of the galaxy. This figure is very instructive. It tells us that if we
want to investigate the metallicity gradient in galaxies, there is no point in
considering these inner regions, since here almost certainly the
SSP-approximation is unjustified, so that the luminosity-weighted metallicity
is probably very uncertain. To verify this, in Fig.\ref{Grads} we plot the
metallicity gradients of Kuntschner et al. (2010) against the Spitzer [3.6] --
[4.5] color gradients (see Section \ref{spitzercol}). We see a good correlation
for the old galaxies, but for galaxies with young central ages, the correlation
breaks down. Although we have tried in Peletier et al. (2012) not to use the
inner, younger, region to fit the gradients, this approach is difficult, and
tricky (see Fig.~\ref{Kgrad}. On the other hand, the good
correlation in Fig.~\ref{Grads} shows that the [3.6] -- [4.5] color probably has a strong metallicity dependence.

Plotting now these gradients against mass in Fig.\ref{Grads} (right) we see no relation between the two
quantities, except that the gradients for the old galaxies are mostly between 0.05 and 0.10.

\begin{figure}
\begin{center}   
\includegraphics[width=\textwidth,clip]{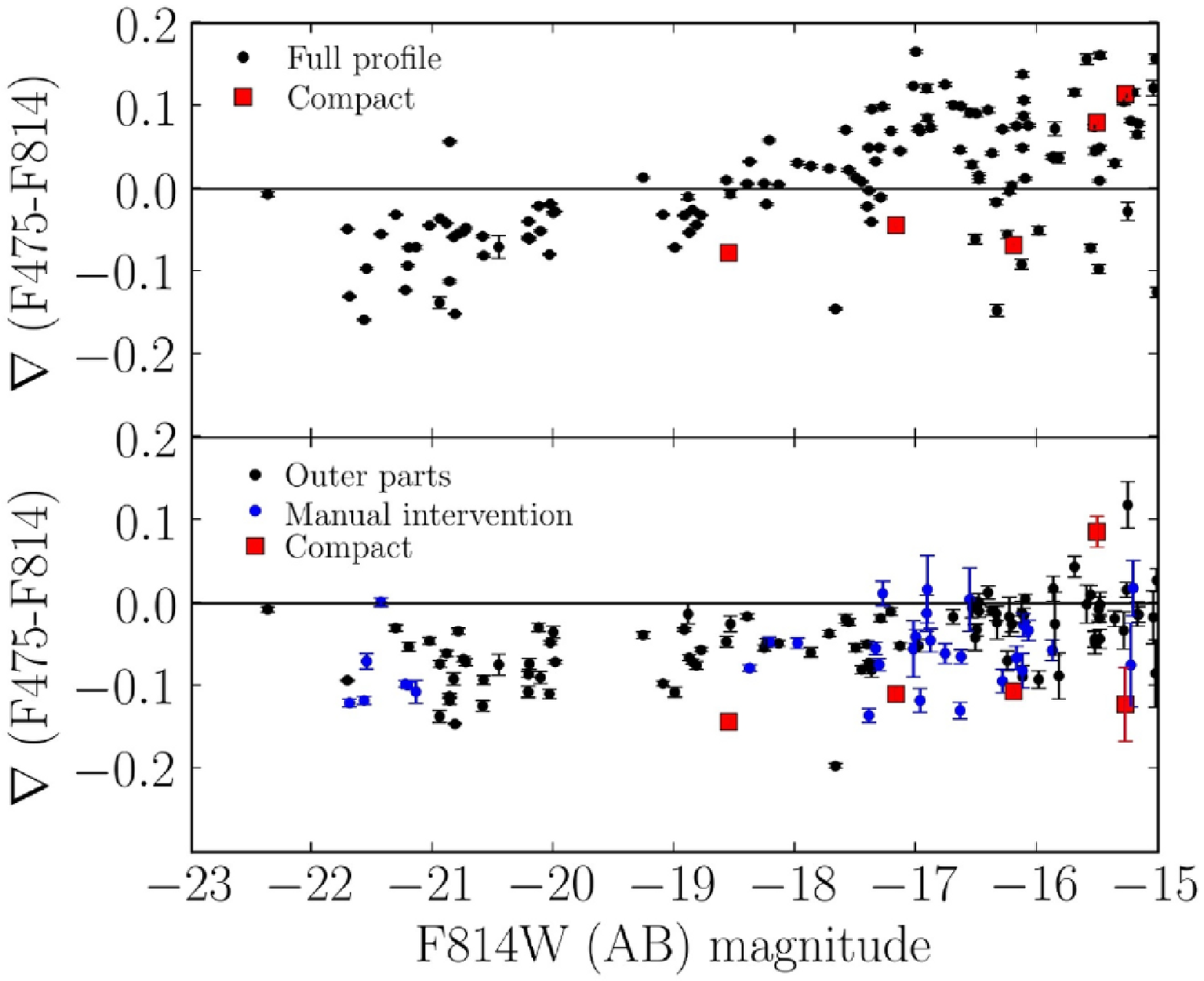}
\label{Gradsig}
\caption{Top: Color gradients of likely Coma members as a function of absolute magnitude. Bottom:
Color gradients of the same galaxies, now excluding the central parts from the fit. Red squares are
compact galaxies from Price et al. (2009). The blue points required manual intervention. (From den Brok et al. 2011).} 
\end{center}
\end{figure}

Realizing this problem with the SSP-based interpretation, we now turn to dwarf ellipticals. Spolaor et al.
(2009) showed a very strong correlation between gradient and mass in the mass range between 3~10$^9$ and
3~10$^{10}$ M$_\odot$, albeit with very few galaxies. Gradients become weaker when going to smaller
galaxies, even becoming positive, indicating positive metallicity gradients. This would have strong
implications on e.g. the effects of galactic SN-driven winds. In den Brok et al. (2011) we have
investigated this effect in detail. We have determined color gradients for a sample of dwarf galaxies in
the Coma cluster using the Coma-ACS survey. In this rich cluster environment the number of young stars in
galaxies is minimized, which means that it is easier to study metallicity gradients. It turns out that the
central regions here are strongly influenced by  small central, younger regions, so-called nuclear
clusters. Since many dwarf ellipticals have nuclear clusters, this effect is very important. In
Fig.~\ref{Gradsig} I show that the gradient -- mass relation before and after removing the nuclear
clusters. In the second case, we indeed find a relation of gradients with mass. Going to smaller galaxies,
the gradients (per dex in radius) become smaller, but remain negative, in agreement with galaxy formation
models. For large galaxies one sees a large scatter in the gradients. Only compact ellipticals do not obey the
gradient - mass relation. They behave in many ways as the central regions of much larger galaxies.

Given the poor reproduction of gradients by simulations, we can only speculate
what these results mean in terms of galaxy formation models. The observed trends
with galaxy mass imply that somehow the mass of the galaxy, or the potential
well are important in shaping the gradient (see also the Mg~b -- v$_{esc}$
relation (Scott et al. 2009). However, given the scatter in gradients at a given
magnitude, this cannot be the only important process. Another clue comes from
the S\'ersic index -- gradient relation. Galaxies with nearly exponential profiles
have flatter gradients than galaxies with higher S\'ersic indices (den Brok et al. 2011). Higher
S\'ersic indices are thought to be the results of processes involving violent
relaxation. Since this is a non-dissipational process by itself, the strong
gradients may point at a history of moist or wet mergers.

\section{Future prospects}
\label{future}

The field of stellar populations is very much alive, and will stay alive for a
long time to go, since it is by far the most accessible way to study galaxies in
the very distant Universe. Led by the Hubble Space Telescope, we know much more
about the Star Formation History of the Universe than 10 years ago. However,  as
far as the interpretation of observations of galaxies at very high redshifts is
concerned,  stellar population modelling lies far behind observations, and it is
very important that we make a huge effort together to catch up. This mainly
involves the modelling in the NUV and FUV, where just a few stars dominate the
integrated stellar populations. Also other areas need work: a lot of effort is
needed in the near-IR, to understand better how stars evolve, so that we can use
near-IR features to study the evolution of galaxies. For a few other challenges,
I refer to the excellent list by Brinchmann (2010).  To these I add the
understanding of abundance ratios in galaxies. We are slowly getting ready to
apply the techniques that have been so powerful for resolved galaxies to integrated spectra. Interpreting abundance ratios will
give us another dimension that will help us to solve the puzzle of galaxy
evolution. Abundance ratios affect everything. The fact that they also affect
broadband colors (Ricciardelli \& Vazdekis 2012) shows that also photometric
redshifts will suffer from systematic errors. Another issue is the IMF, which at
the moment is popular again. Understanding the dark mater in galaxies, and
therefore also the stellar mass, is fundamental. It is still very difficult to
derive accurate stellar masses of galaxies, but we are making progress. The main
difference with the past is we now have new observations from gravitational
lensing which independently can constrain the IMF slope.

In the future there will be many new facilities useful for stellar population studies. A few
examples, ranked by increasing telescope size, are X-shooter on the VLT, the
JWST and the E-ELT. They will give the high spectral resolution and the high
signal-to-noise needed to derive accurate star formation histories which can be
combined with high resolution LOSVDs. They will open new fields, such as the
redshifted optical and near-IR. At the same time GAIA will determine the stellar
populations in our Galaxy in much more detail as before, so that we can use the
Milky Way as template for studies of other galaxies.

\begin{figure}
\begin{center}   
\includegraphics[height=6cm,clip]{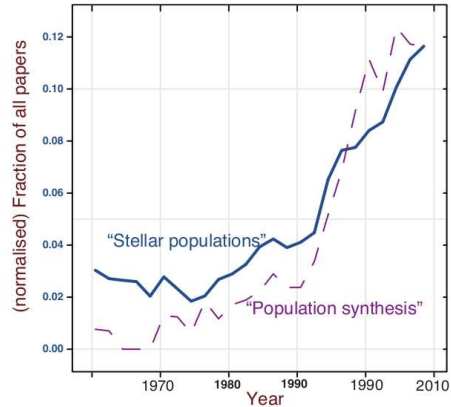} 
\caption{The fraction of papers in A\&A, AJ, ApJ and MNRAS each year that
mention {\it Stellar populations} (solid line) or {\it Population synthesis} (dashed line, scaled up by a
factor of 8) in their abstracts. Currently about 12\% of all papers mention {\it Stellar populations}.
(from Brinchmann 2010).}
\label{popularity}
\end{center}
\end{figure}

The fact that more and more data is available of always higher quality is also reflected 
in an increasing popularity of the field of stellar populations within the astronomical community
(see Fig.~\ref{popularity}).
Brinchmann (2010) shows  that many
astronomers feel interested in helping to solve the problems in the field of stellar populations. 
I am sure that a lot of progress will come from you, the attendants of the
Winter School.

I would like to finish this review with a quote of Renzini (2006):  {\it Baryon
physics, including star formation, black hole formation and their feedbacks, is
highly nonlinear, and it is no surprise if modeling of galaxy evolution relies
heavily on many heuristic algorithms, their parameterization, and trials and
errors. Dark matter physics, on the contrary, is extremely simple by comparison.
Thus, the vindication of the CDM paradigm should be found in observations
demonstrating that the biggest, most massive galaxies are the first to disappear
when going to higher and higher redshifts. This is indeed what has not been seen
yet, and actually there may be hints for the contrary. }  This statement is
probably supported by Tolstoy et al. (2009): {\it  The hierarchical theory of
galaxy formation contains at its heart the concept of smaller systems
continuously merging to form larger ones. This leads to the general expectation
that the properties of the smaller systems will be reflected in the larger.
(...) From recent abundance studies of low-metallicity stars in dSphs, it seems
likely that there exist only narrow windows of opportunity when the merging of
dwarf galaxies to form larger systems would not lead to inconsistencies.} The
biggest challenge for the future is to connect baryon and dark matter physics. But finding the
solution here obviously is not so easy. 

\section*{Acknowledgments}
I thank Johan Knapen and Jes\'us Falc\'on-Barroso for having organized a very interesting and pleasant
meeting, and the IAC secretaries for making sure that everything ran smoothly.

%

%% file: chapter_peletier/bmacros.tex
\def\bea{\begin{eqnarray}}
\def\eea{\end{eqnarray}}
\def\deffn#1{{\bf#1}}
\arraycolsep=1pt

\def\ev{{\,\rm eV}}
\def\kev{{\,\rm keV}}
\def\mev{{\,\rm MeV}}
\def\gev{{\,\rm GeV}}
\def\Hz{{\,\rm Hz}}\def\MHz{{\,\rm MHz}}\def\GHz{{\,\rm GHz}}\def\THz{{\,\rm THz}}
\def\m{{\,\rm m}}\def\cm{{\,\rm cm}}\def\nm{{\,\rm nm}}
\def\N{\,{\rm N}}
\def\gm{{\,\rm gm}}\def\kg{{\,\rm kg}}
\def\T{\,{\rm T}}
\def\J{\,{\rm J}}
\def\pc{\,{\rm pc}}
\def\kpc{{\rm\,kpc}}
\def\Mpc{{\rm\,Mpc}}
\def\msun{\,{\cal M}_\odot}
\def\kms{\,{\rm km}\sec^{-1}}
\def\yr{{\,\rm yr}}
\def\Myr{{\rm\,Myr}}
\def\Gyr{{\rm\,Gyr}}
\def\del{\vec\nabla}
\def\Ht{H$_2$}
\def\twelveco{$^{12}\hbox{C}\hbox{O}$}
\def\thirteenco{$^{13}\hbox{C}\hbox{O}$}
\def\msun{\,{\rm M}_\odot}
\def\muB{\mu_{\rm B}}\def\muN{\mu_{\rm p}}
\def\rhs{\hbox{\caps rhs}}
\def\lhs{\hbox{\caps lhs}}
\def\rms{{\caps rms}}
\def\mcmc{{\caps mcmc}}
\def\df{{\caps df}}
\def\Omegap{\Omega_{\rm p}}
\def\Rg{R_{\rm g}}
\def\cF{{\cal F}}\def\cK{{\cal K}}\def\cJ{{\cal J}}
\newcommand {\hi} {{\rm H}\,{\small\rm I}}
\def\d{{\rm d}}\def\e{{\rm e}}
\def\i{\relax\ifmmode{\rm i}\else\char16\fi}
\def\p{\partial}
\def\deg{^\circ}

{\newif\ifnotend
\notendtrue
\def\veclist{ABCDEFGHIJKLMNOPQRSTUVWXYZabcdefghijklmnopqrstuvwxyz.}
\def\top#1#2.{#1}
\def\tail#1#2.{#2.}
\loop\expandafter\xdef\csname v\expandafter\top\veclist\endcsname%
{{\noexpand\bf\expandafter\top\veclist}}
\edef\veclist{\expandafter\tail\veclist}
\if\veclist.\notendfalse\fi\ifnotend\repeat}

%
%
\def\real{\Re\hbox{{\eightrm e}}}                   
\def\imag{\Im\hbox{{\eightrm m}}}                   
%
%
\def\fracj#1#2{{\textstyle{#1\over#2}}}
%
\def\spose#1{\hbox to 0pt{#1\hss}}
\def\lta{\mathrel{\spose{\lower 3pt\hbox{$\mathchar"218$}}
     \raise 2.0pt\hbox{$\mathchar"13C$}}}
\def\gta{\mathrel{\spose{\lower 3pt\hbox{$\mathchar"218$}}
     \raise 2.0pt\hbox{$\mathchar"13E$}}}

%
\font\gkvecten=cmmib10
\font\gkvecseven=cmmib7
\let\boldgrk=\gkvecten
\let\boldgrksc=\gkvecseven

\def\gkthing#1{{\mathchoice%
	{\hbox{{\boldgrk\char#1}}}
	{\hbox{{\boldgrk\char#1}}}
	{\hbox{{\boldgrksc\char#1}}}
	{\hbox{{\boldgrksc\char#1}}}}}

\def\valpha{\gkthing{11}}
\def\vbeta{\gkthing{12}}
\def\vgamma{\gkthing{13}}
\def\vdelta{\gkthing{14}}
\def\vepsilon{\gkthing{15}}
\def\vzeta{\gkthing{16}}
\def\veta{\gkthing{17}}
\def\vtheta{\gkthing{18}}
\def\viotaeta{\gkthing{19}}
\def\vkappa{\gkthing{20}}
\def\vlambda{\gkthing{21}}
\def\vmu{\gkthing{22}}
\def\vnu{\gkthing{23}}
\def\vxi{\gkthing{24}}
\def\vpi{\gkthing{25}}
\def\vrho{\gkthing{26}}
\def\vsigma{\gkthing{27}}
\def\vtau{\gkthing{28}}
\def\vupsilon{\gkthing{29}}
\def\vphi{\gkthing{30}}
\def\vchi{\gkthing{31}}
\def\vpsi{\gkthing{32}}
\def\vomega{\gkthing{33}}

\def\vnabla{{\mathchoice
{\hbox{{\boldsym\char114}}}
{\hbox{{\boldsym\char114}}}
{\hbox{{\boldsymsc\char114}}}
{\hbox{{\boldsymsc\char114}}}
}}%

\def\vDelta{{\bf\Delta}}
\def\vOmega{{\bf\Omega}}

\def\figref#1{Fig.~\ref{#1}}

\def\Phieff{\Phi_{\rm eff}}

\def\const{\hbox{constant}}

\renewcommand{\[}{\begin{equation}}\renewcommand{\]}{\end{equation}}